\documentclass[traditabstract]{aa}

\newcommand{\elodie}{{\tiny ELODIE}}

\newcommand{\teff}  {\mbox{T$_\mathrm{eff}$}}
\newcommand{\feh}   {\mbox{[Fe/H]}}
\newcommand{\mh}   {\mbox{[M/H]}}
\newcommand{\logg}  {\mbox{$\log$ g}}

\newcommand{\kms}{km~s$^{-1}$}

\usepackage{graphicx,amssymb,amsmath,natbib,hyperref,multirow}
\usepackage[usenames,dvipsnames]{xcolor}
\usepackage{rotating}
\usepackage{txfonts}

\definecolor{mygreen}{rgb}{0.13,0.73,0.13}
\newcommand{\lc}[1]{\textit{\textcolor{mygreen}{N1}}}

\begin{document}
\title{Solar twins in the ELODIE archive 
\thanks{Based on data extracted from the \elodie\ archive at Observatoire de Haute--Provence (OHP), http://atlas.obs-hp.fr/elodie/ }
}

 \author{D. Mahdi\inst{1}
 \and C. Soubiran\inst{1}
 \and S. Blanco-Cuaresma\inst{2}
 \and L. Chemin\inst{1}
  }

\institute{LAB UMR 5804, Univ. Bordeaux  - CNRS, F-33270, Floirac, France\\
             \email{soubiran@obs.u-bordeaux1.fr}
 \and Observatoire de Gen\`eve, Universit\'e de Gen\`eve, CH-1290 Versoix, Switzerland }           

 \date{Received \today; accepted XX}

  \abstract{
     A large dataset of $\sim$2800 spectra extracted from the \elodie\ archive was analysed in order to find solar twins. A list of stellar spectra closely resembling  the spectrum of the Sun was selected by applying a purely differential method, directly on the fluxes. As solar reference, 18 spectra of asteroids, Moon and blue sky were used. Atmospheric parameters and differential abundances of 8 chemical elements were determined for the solar twin candidates, after a careful selection of appropriate lines. The Li feature of the targets was investigated and additional information on absolute magnitude and age was gathered from the literature. HIP076114 (HD138573) is our best twin candidate, looking exactly like the Sun in all these properties. 
  }



   \maketitle
%

\section{Introduction}
\label{sec:intro}
The definition of solar twin was first introduced by \cite{1981A&A....94....1C}. Solar twins are stars which have the same physical properties as those of the Sun : mass, radius, luminosity, chemical composition, rotation, activity. Consequently, the spectrum of a solar twin should be identical to that of the Sun.  A way to find solar twins is thus to compare stellar spectra to solar spectra, and to identify those looking the most similar. There are several motivations to search for solar twins. The Sun, as the best known star, is used as the fundamental standard in many astronomical calibrations. A motivation to identify stars that replicate the solar astrophysical properties is the need to have other reference stars, observable during the night under the same conditions as any other target. For example, \cite{2010A&A...512A..54C} used a set of solar twins to calibrate the effective temperature scale from the infrared flux method, and showed that this model independent technique is not affected by systematics \citep{2014MNRAS.439.2060C}. Sun-like stars can also help us to understand whether the chemical composition of the Sun compared to other stars is unusual \citep{2009A&A...508L..17R, 2015A&A...579A..52N} and to explore the connections between planet formation and stellar chemical composition \citep{2014A&A...572A..48R}. Solar twins represent natural candidates for harboring planetary systems similar to ours. Finally solar twins may also give some clues on where and how the Sun formed in our Galaxy. Such stars may have formed from the same conditions as the Sun and from the same molecular cloud. In that case, one refers to solar siblings which are expected to also share the same kinematical properties as the Sun \citep{2015A&A...575A..51L}.  
    
 Since the pioneering work by \cite{1978A&A....63..383H}, the hunt of the closest solar twins has been very active. Giusa Cayrel de Strobel and her collaborators made a huge contribution to the subject with the detailed spectroscopic analysis of many candidates \citep{1981A&A....94....1C, 1989A&A...211..324C, 1993A&A...274..825F}, and a review of the state of the art \citep{1996A&ARv...7..243C}.  More recently, a large scale search of solar twins has been performed  by \cite{2012MNRAS.426..484D, 2014MNRAS.439.1028D, 2015A&A...574A.124D} using the {\tiny ESO  FEROS} and {\tiny HARPS} archives, with 17 twins identified. \cite{2014A&A...563A..52P} conducted a spectroscopic survey of solar twin stars within 50 parsecs of the Sun, and identified some candidates on the basis of photometric colours and atmospheric parameters determined from high resolution spectra. Several other studies have used the differential analysis of high resolution spectra to identify solar twins but very few stars were found very similar to the Sun. HIP079672 (HD146233 - 18 Sco) is the most studied one, and often claimed to be the best one \citep{1997ApJ...482L..89P, 2004A&A...418.1089S, 2006ApJ...641L.133M, 2009PASJ...61..471T}, but it was finally found not to be a perfect twin, with slighlty higher effective temperature and surface gravity, different chemical pattern \citep{2014ApJ...791...14M}, higher mass and younger age \citep{2011A&A...526L...4B, 2012A&A...546A..83L}. HIP114328 (HD218544) is considered by \cite{2014A&A...567L...3M} as an excellent solar twin to host a rocky planet due to its abundance pattern very similar to that of the Sun, despite a lower Li content and indications of being older. The abundance pattern of HIP102152 (HD197027) was claimed to be the most similar to solar of any known solar twin by \cite{2013ApJ...774L..32M}, while HIP056948 (HD101346) is considered as a prime target in the quest for other Earths due to its similarity to the Sun \citep{2012A&A...543A..29M, 2009PASJ...61..471T, 2007ApJ...669L..89M}. In fact, most of the solar twins are found to have chemical compositions different from that of the Sun when submitted to high precision differential analysis, suggesting that the Sun has an unusual abundance pattern \citep{2015A&A...579A..52N}.  Interestingly \cite{2011A&A...528A..85O} claimed that M67-1194 is the first solar twin known to belong to a stellar association. The chemical similarity between the Sun and M67-1194 suggests that the Sun once formed in a cluster like M67. However  \cite{2012AJ....143...73P} demonstrated with dynamical arguments that M67 could not have been the birth cluster of the Sun and also excluded the possibility that the Sun and M67 were born in the same molecular cloud. 
     
     In this paper we searched for new solar twins in a large  sample of spectra selected from the \elodie\ archive \citep{2004PASP..116..693M}, using  18 spectra of asteroids, the Moon and the day sky as solar reference. This observational material is described in Sect. \ref{s:spec}. We proceeded in several steps. First, as described in Sect. \ref{s:tgmet}, twin candidates were identified with a minimum distance method  already used and validated for that task by \cite{2004A&A...418.1089S}. Then their atmospheric parameters (effective temperature \teff, surface gravity \logg, metallicity \mh) were derived and compared to those of the Sun (Sect. \ref{s:ap}). We inspected the spectral range around the Li lines at 670.78 nm to find twin candidates showing the same Li depletion as the Sun (Sect. \ref{s:li}). In Sect. \ref{s:ab} we explain how we selected the good lines from which abundances where determined differentially to the solar spectra.  Finally we searched for extra information on the targets (Sect. \ref{s:xhip}) and discuss our findings in Sect. \ref{s:discuss}.
     
\section{\elodie\ spectra}
\label{s:spec}
All the spectra used in this paper were retrieved from the archive of the \elodie\ echelle spectrograph \citep{2004PASP..116..693M}. \elodie\ was on the  1.93 m telescope at Observatoire de Haute--Provence (OHP) between late 1993 and mid 2006. It was designed for very precise radial velocity measurements  \citep[the discovery of the first extra-solar planet 51 Peg B by][was made with this instrument]{1995Natur.378..355M} but it has also been used for many other programs in stellar physics and galactic structure.  The spectra cover the interval from 389.5 - 681.5 nm and are recorded as 67 orders with a resolution of R$\simeq$42\,000 \citep{ 1996A&AS..119..373B}. The archive provides the spectra in two formats. In the {\it S2D} format, the spectra are recorded as 67 $\times$ 1024 pixels with the coefficients of the pixel to wavelength relation of each order being stored in the FITS header, and deblazed (the blaze function being stored in a FITS extension).  In the {\it spec} format, the orders are reconnected and resampled in wavelength with a constant step of 0.05 \AA, covering the range 400-680 nm (the first 3 blue orders are not included as their signal to noise ratio is usually very low). There are several modes of observation related to  the two sets of optical twin-fibers, direct and scrambled : one fiber is assigned to the object while the second one can be masked, or on the sky, or illuminated by a thorium-argon lamp. The \elodie\ archive also releases the radial velocities calculated at the telescope by cross-correlation of the spectra with numerical masks of spectral type F0 or K0.  The gaussian fit of a cross-correlation function (CCF) provides the fitted radial velocity, the full width at half maximum (FWHM) and the  amplitude of the correlation peak. In the case of spectroscopic binaries, a double gaussian is fitted. The CCFs are provided as FITS files, with the parameters of the gaussian fit in the header. More detailed information on the \elodie\ archive data products is available in the  on-line user's guide\footnote{http://atlas.obs-hp.fr/elodie/intro.html }.

The \elodie\ archive contains more than 35\,000 public spectra and there are $\sim$8\,000 distinct object names (with possible duplication due to unresolved aliases).  For this work on solar twins, only spectra with a signal to noise ratio (S/N) at 550 nm  greater than 70, a measured radial velocity and an identification resolved by Simbad were selected. Spectra showing an enlarged CCF with FWHM greater than 12 \kms\ or a double-peaked CCF  were rejected. No other selection based on colour or spectral type was done. Large series of observations of a single star were shortened by retrieving from the archive only one observation per night and five observations in total  (the ones with the highest S/N, favouring exposures without simultaneous thorium-argon). Repeated observations of the same stars were used to test the consistency of our methods and to provide more robust results, as explained in Sect. \ref{s:ap} and Sect. \ref{s:ab}. The selection corresponds to  2\,784 spectra of 1\,165 different stars which were downloaded as FITS files in {\it S2D} and {\it spec} formats, together with the CCF files. In addition, 18 spectra of solar system bodies and  day sky light were retrieved to be used as reference solar spectra. These solar spectra were obtained at different dates, with various S/N and fiber modes giving a representive range of observing conditions for the other stars. The list of solar spectra is presented in Table \ref{t:sun_spectra}.

\begin{table}[h]
  \centering 
  \caption{Solar spectra available in the ELODIE archive. The spectra can easily be retrieved from the archive using the date of observation and the exposure serial number (imanum). S/N is provided in the header of the spectrum, for the order 46 centered on 550 nm. The FWHM is that of the CCF retrieved from the archive. The exposure type indicates whether the observation was made with a simultaneous thorium calibration (OBTH) or not (OBJO) and with direct or scrambled fibers (d and s respectively, after 1997). }
  \label{t:sun_spectra}
\begin{tabular}{lcccl}
\hline
Name     & Date /Imanum &  S/N & FWHM & Exposure \\
              &                              &  550 nm & \kms & type \\
\hline
CERES  &  19950206 /0021  & 107.3  &  11.13 & OBTH\\ 
CERES  &  20000327 /0023  &  117.3 &  11.01 & OBJOd \\  
CERES  &  20040209 /0007  & 130.8 & 11.03& OBJOd\\ 
Sky &  19960629 /0007 & 335.6& 11.07 & OBJO \\
Sky   & 20060613 /0010 &190.9 &10.96 & OBJOs \\
Sky  &  20060613 /0011 &286.3 &10.98 & OBJOs\\
Sky & 20021127 /0009 & 430.1 & 10.95 & OBJOd \\
Sky   & 20060601 /0027    & 142.4& 11.04 & OBJOd\\
MOON & 19980114 /0008 & 381.4 & 11.06 & OBJOd \\
MOON & 19981001 /0012 & 110.8 & 10.99 & OBTHs \\
MOON & 19991222 /0013 & 139.6 & 11.05 & OBJOd \\
MOON & 19991222 /0014 & 156.5 & 11.06 & OBJOd \\
MOON & 20000124 /0028 & 200.0 & 11.0 5& OBJOd \\
MOON & 20000124 /0029 & 224.9 & 11.05 & OBJOd \\
MOON & 20000609 /0008 & 350.8 & 11.08 & OBJOd \\
MOON & 20000609 /0009 & 246.4 & 11.08 & OBJOd \\
VESTA & 19950110 /0020 & 98.7 & 11.11 & OBTH \\
CALLISTO & 19990813 / 0037 & 218.2  & 11.10 & OBJOd \\
\hline
\end{tabular}
\end{table}

\section{Minimum distance between spectra}
\label{s:tgmet}
To measure the degree of similarity between two spectra, of a target and of the Sun, we applied the TGMET code developed by \cite{1998A&A...338..151K}, a purely differential method implementing a minimum distance criterion. TGMET was already applied by \cite{2004A&A...418.1089S} on a selection of $\sim$ 200 \elodie\ G dwarf spectra in order to identify solar twins. 
 
 The method was applied to {\it S2D} spectra,  on orders 21 to 67, corresponding to the wavelength range 440--680 {\AA}, with rejection of the under-illuminated 
edges of the orders. The order 63 ($\sim$ 648 -- 652.5 nm) was rejected because it is very affected by telluric lines. We briefly recall the principle of the TGMET method. Two spectra to be compared need to be put on the same scale, in wavelength and in flux. The wavelength alignement is made by shifting the compared spectrum to the radial velocity of the solar spectrum and
by resampling it to the same wavelength points. This operation
implies an interpolation between wavelengths for the compared spectrum, which is performed with the quadratic Bessel formula. Once the horizontal ajustement is done, the vertical one is perfomed by fitting the flux of the compared spectrum to the solar spectrum. To do so, a simple factor is determined by least-squares (the two stars having roughly the same temperature it is not necessary to introduce a slope).  The reduced 
$\chi^2$  of that fit, computed order by order over nearly 40\,000 wavelengths in total, is thus a distance quantifying the similarity between the two spectra. 
However following \cite{1998A&A...338..151K}, we did not adopt the real reduced  $\chi^2$ for the distance between two spectra, which would imply taking into account the noise on each pixel. Instead, we computed an averaged and normalized instrumental response curve that we used as the  weighting function of the fit. This smooth function reflects the global variation of S/N over each order (the edges of an order are under-exposed compared to its central part). It gives a similar weight to the continuum and to the wings and bottom of absorption lines, 
contrary to a weighting function based on the photon noise. Several tests made in \cite{1998A&A...338..151K} demonstrated its higher performance, especially at high S/N. 

TGMET was applied to compute the distance of  each solar spectrum to each of the other spectra. The TGMET output for a given solar spectrum  resulted in 2\,801 distances (for 2\,784 target spectra plus the 17 other solar spectra) that were sorted by increasing value.  Since solar spectra were processed like the other targets, they were used to verify the results : the nearest neighbours of a given solar spectrum are expected to always be the other solar spectra. This is verified in most cases, but sometimes some stars are found closer to a given solar spectrum than other solar spectra, demonstrating that the observing conditions impact the results. This also means that such stars have spectra very similar to that of the Sun, and thus they are good candidates for solar twins. TGMET was also  run with spectra convolved to a common broadening of FWHM=12\kms\ in order to  minimize the possible effect of a different rotation and macroturbulence in the targets and in the Sun. This however did not produce significantly different results. Other TGMET runs were performed on a smaller number of orders, those identified to carry most information. One run was performed without 10 orders found to give systematically larger distances than the other orders, and more scattered results, among the solar spectra. Another run was performed using only the order 39 centered on the Mg triplet at $\sim$ 520 nm, which is sensitive to effective temperature, surface gravity and metallicity and thus a good region to test the spectrocopic similarity of stars.  The best twin candidates were identified by mixing these results obtained with TGMET runs in different configurations. In total 108 TGMET output files were obtained, corresponding to 18 solar spectra x 2 (convolved and non convolved spectra) x 3 (all orders, selected orders, order 39). The list of solar twin candidates was built by selecting all stars ranked at a smaller distance than the last solar spectrum in these different result files. This gives 56 stars (225 spectra) among which remarkable stars are to be noted. HIP079672 (HD 146233, 18 Sco) is by far the most frequent first neighbour. HIP018413 (HD024409 ), HIP076114 (HD138573), HIP089474 (HD168009) appear in the 5 nearest neighbours in more than 70\% of the result files, followed by HIP118162 (HD224465)  and HIP041484 (HD071148).  The status of these stars as solar twins will be discussed in Sect. \ref{s:discuss}.

\section{Atmospheric parameters}
\label{s:ap}
The atmospheric parameters and abundances of the solar and target spectra were derived with iSpec \citep{2014A&A...569A.111B}, following the recipes described in \cite{2015A&A...577A..47B}.  For that task we  used the \elodie\ spectra in {\it spec} format (i.e. with the orders reconnected and the wavelengths resampled at a constant step) in the range 480-670 nm where iSpec was extensively tested and validated. As  model atmospheres we used MARCS  \citep{2008A&A...486..951G}. 

For the atmospheric parameters, the line list established for the Gaia ESO survey \citep{2015PhyS...90e4010H} was used. It consists of $\sim$ 140\,000 atomic lines extracted from the VALD database among which $\sim$2000 lines of 35 elements have been revisited and evaluated on the basis of the quality of the atomic data and of the spectral synthesis for the Sun and Arcturus. For our use  the recommended lines  were selected, those flagged to be the most reliable ones.   \teff, \logg\ and \mh\ were determined automatically and iteratively. The different steps are :  (1) the nomalization of the spectra by two degree splines at every nanometer on pre-selected points, (2) a first guess of atmospheric parameters by comparing the wings of the \mbox{H$\alpha$} and \mbox{H$\beta$} lines and the Mg triplet to a small pre-computed synthetic grid, (3) the gaussian fit of lines with rejection of badly fitted lines (blended lines, gaps, cosmics, too faint lines,..), (4)  the synthesis of the remaining lines allowing the determination of the atmospheric parameters by least squares. Vmic and Vmac were set as free parameters while $v\sin i$ was set to 2 \kms\ for all the targets. These three parameters are possibly degenerated but the selection of spectra with FWHM $\le$ 12 \kms\ prevented us from dealing with stars rotating much faster than the Sun or with higher macroturbulent velocity.  In practice  615 to 805 lines of 28 elements were used for the determination of atmospheric parameters, depending on the spectrum.

In order to evaluate systematic errors and correct them, we investigated in  detail the results obtained for the 18 solar spectra. A very high consistency was obtained. For \teff\ and \logg, the averaged  values are  5773K and 4.32 with standard deviations of 9 K and 0.02 dex respectively. The \teff\ difference with the fundamental value, 5777 K, is negligible. It is worth to note that the fundamental Teff of the Sun was recently revised by \cite{2015arXiv150606095H} to be \teff$_\odot$ = 5771 $\pm$ 1 K , putting our determination even closer to that fundamental value. We found a systematic shift of -0.12 dex for gravity  with respect to the fundamental value of \logg$_\odot$ = 4.44. This bias was found in previous works based on iSpec \citep{2015A&A...577A..47B}. A correction of +0.12 dex was thus applied for the rest of the analyzed targets, establishing a global zero point centered on the solar reference parameters.  Similarly, we found an average metallicity [M/H]=-0.15 dex for the solar spectra, determined from lines of various elements. The global metallicities of the targets, determined from the same lines, were thus corrected by this value to be relative to the Sun. 

As stated in the \elodie\ user's manual, it is suspected that for image types OBTH, the presence of the thorium-argon orders interleaved with the stellar orders may lead to pollution of the stellar spectrum by highly saturated Argon lines which can contaminate the adjacent stellar orders. So in principle this kind of spectra should not be used for the determination of atmospheric parameters and abundances. However we obtained similar atmospheric parameters for the Sun when we averaged them from the 15 OBJ and 3 OBTH spectra separately : negligeable differences of 8K, 0.02 dex and 0.01 dex were measured in \teff, \logg\ and \mh\ respectively. There is thus apparently no impact of the image type on the atmospheric parameters, probably due to the selection of well fitted lines and to the synthesis method.

In the target sample, 48 stars have dupplicate observations from which we estimated the internal consistency of the derived atmospheric parameters. For those stars the standard deviation around the mean Teff ranges from 1 K to 20K, with a median of 9K. For \logg\ the median standard deviation is 0.02 dex, with no higher value than 0.04. For \mh\ all the stars have a median standard deviation of 0.01 dex, except HIP097336 which has a standard deviation of 0.02 dex for 5 spectra. Although these values show the excellent consistency of the iSpec determinations of atmospheric parameters from one spectrum to another, we still verified that the 16 stars observed with both modes OBJ and OBTH have derived parameters in good agreement from both types of exposures. 


HIP079672 (18 Sco) is one of the Gaia benchmark stars recently studied by \cite{2015arXiv150606095H} who determined fundamental \teff=5810 K and \logg=4.44 from the defining relations,  independently of spectroscopy. The Gaia benchmark stars are intended to serve as calibrators for spectroscopy and thus can be used to evaluate our determinations. Only one spectrum is available for HIP079672 in the \elodie\ archive from which we determined \teff= 5793$\pm$9 K and \logg=4.43$\pm$0.015, in good agreement with the fundamental values. We thus confirm that HIP079672  is slightly hotter than the Sun. 

\begin{figure}[h!]
\begin{center}
 \includegraphics [width=6 cm] {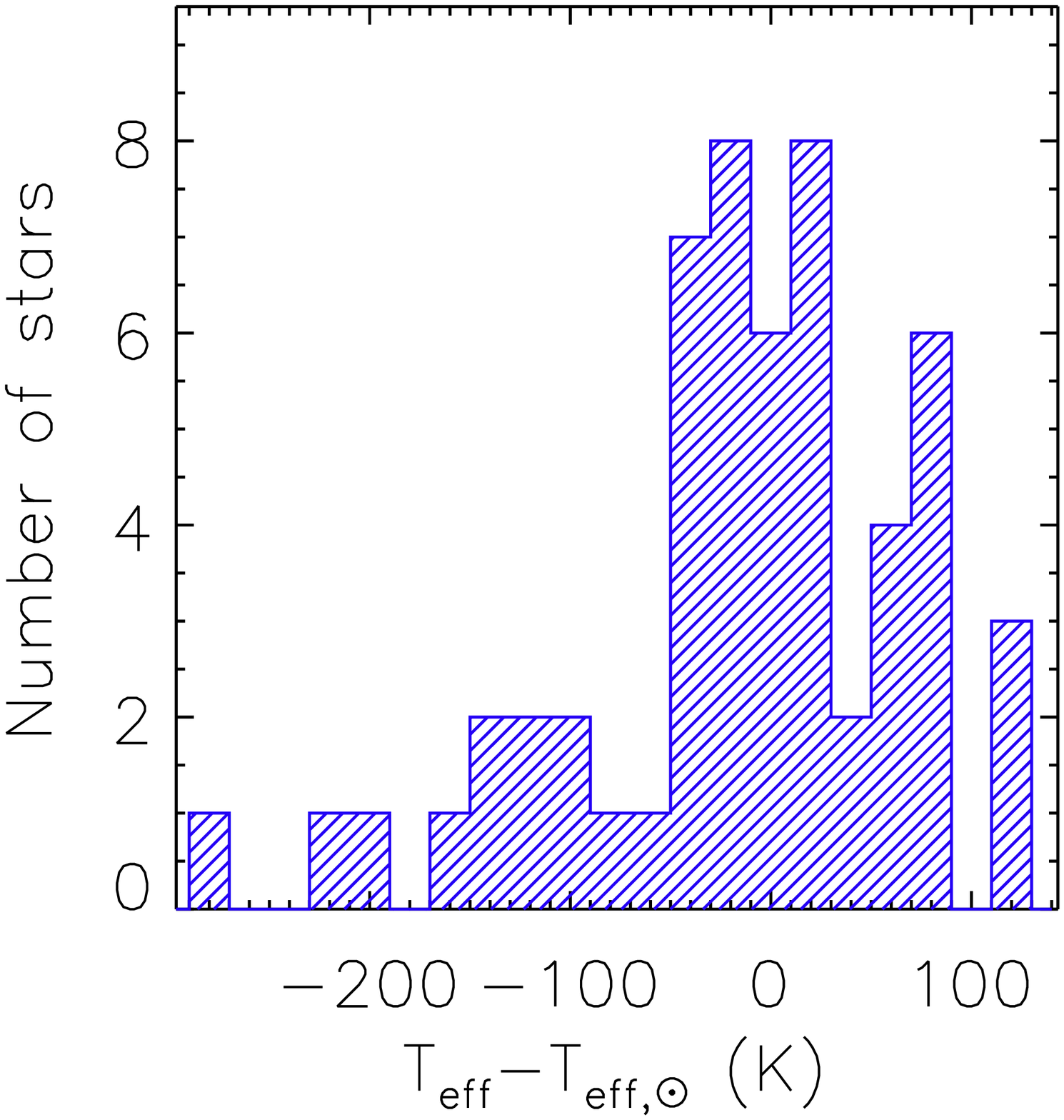} 
\includegraphics [width=6 cm] {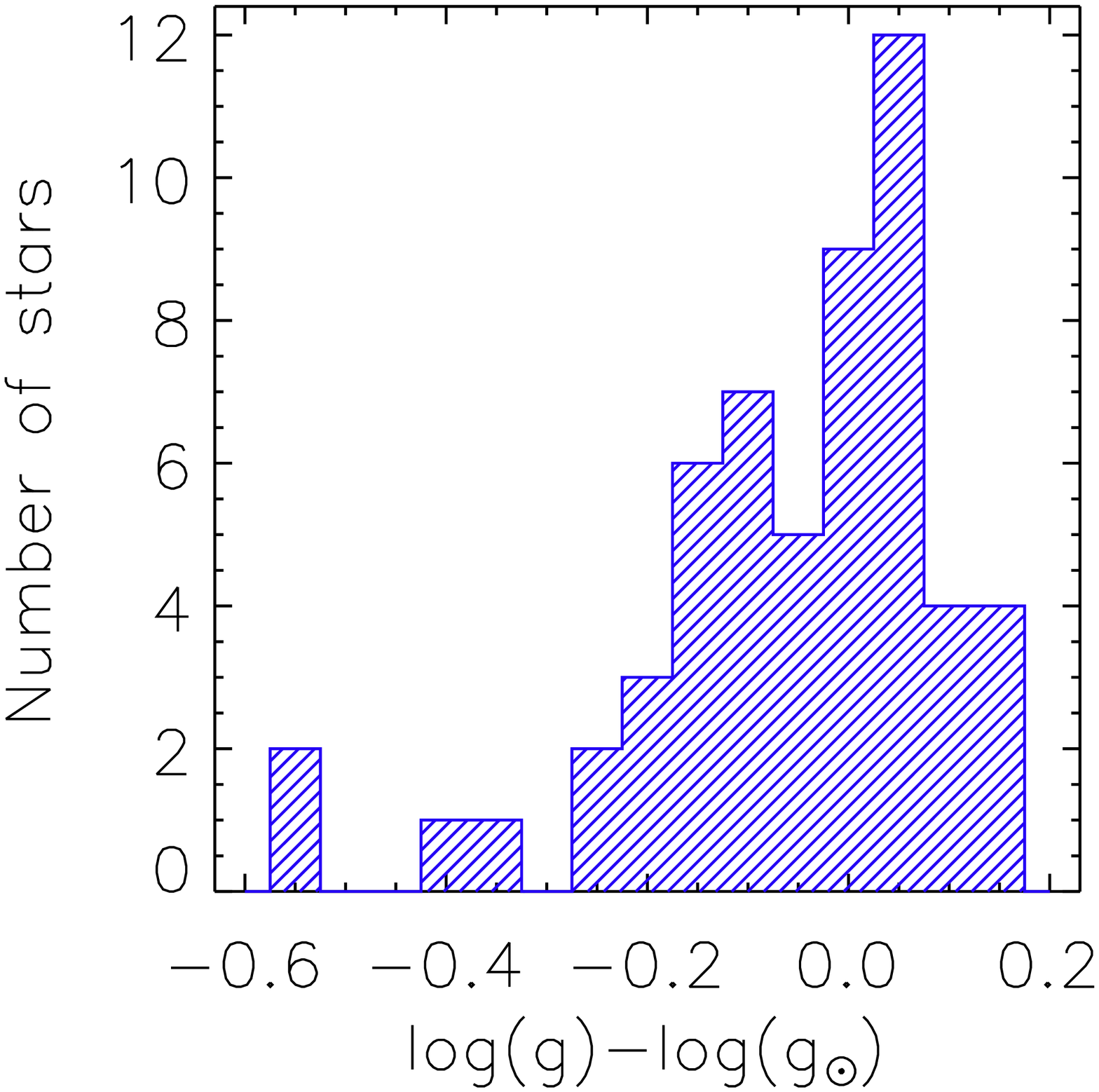}
\includegraphics [width=6 cm] {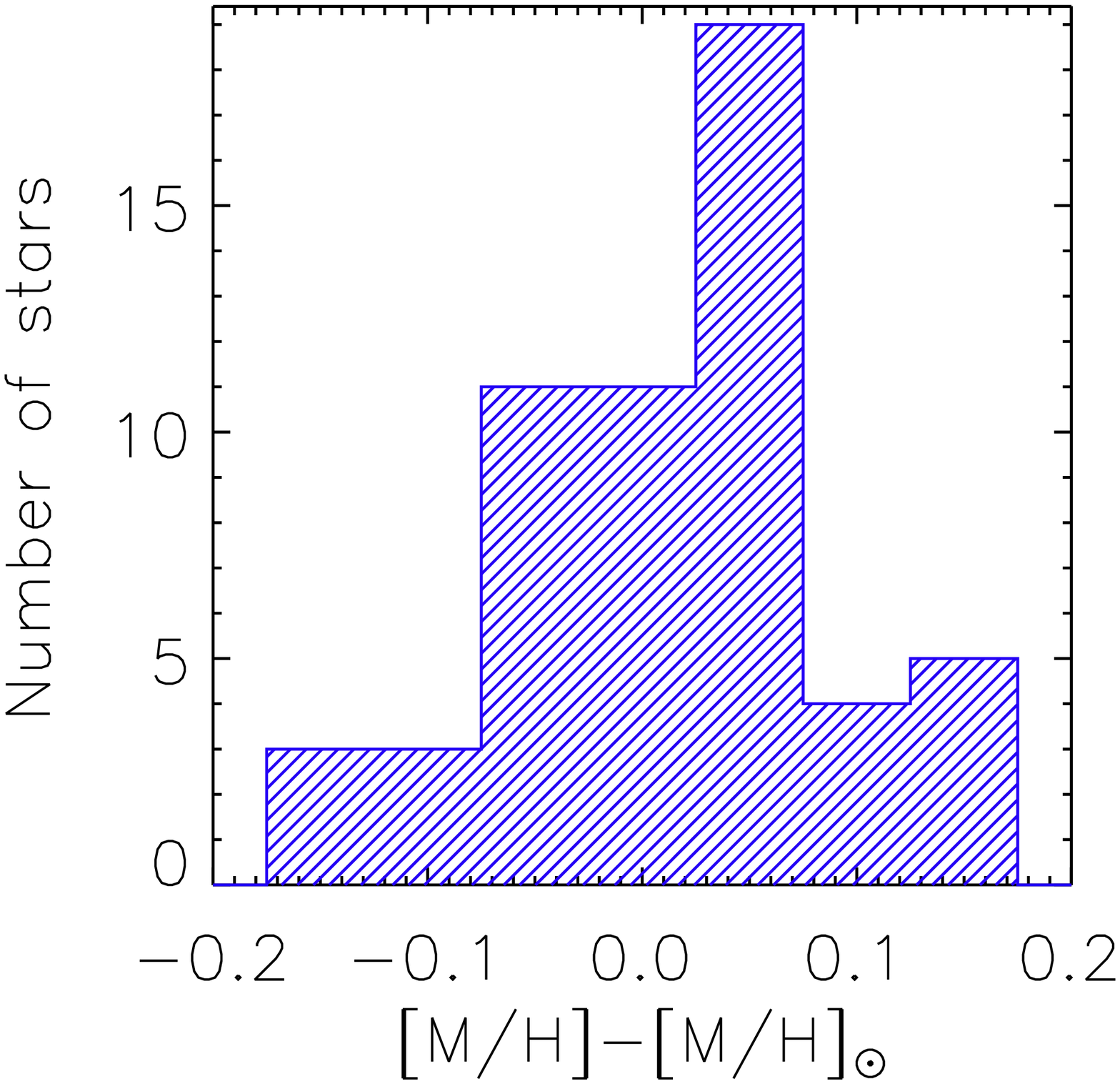}
\caption{Histograms of the differences between the atmospheric parameters of the twin candidates and of the Sun, as obtained by iSpec.}
\label{f:ap}
\end{center}
\end{figure}

Atmospheric parameters are presented in Table \ref{t:ap}. The histograms of the differences with the solar values are presented in Fig. \ref{f:ap}. There are 22 stars which fall into the category of solar twin according to the criteria adopted by \cite{2014A&A...567L...3M} : they differ by less than 100 K in \teff, 0.1 dex in \logg, and 0.1 dex in metallicity from the Sun values. We flag such stars with AP category = B in Table \ref{t:ap}. Three remarkable stars have to be noted as they  have the same atmospheric parameters as the Sun, within 25K in \teff\ and 0.05 dex in \logg\ and \mh\ (flagged with AP category = A in Table \ref{t:ap}) : HIP076114, HIP085244 and HIP088194.  The status of these stars as solar twins will be discussed in Sect. \ref{s:discuss}.

Interestingly some stars have quite different parameters from the Sun (flagged with AP category = C in Table \ref{t:ap}). For instance HIP065721 is a sub-giant cooler than the Sun and slightly more metal-poor. This example illustrates the degeneracy of atmospheric parameters at the resolution of 42\,000 when the overall spectrum is considered.  This suggests that the spectrocopic comparison between spectra should be performed differentially, on a line by line basis.

\begin{table*}[h]
 \centering 
 \caption{Atmospheric parameters obtained with iSpec. \logg\ and \mh\ were scaled to the solar values. In the fifth column, N is the number of spectra available for the star, from which the averaged atmospheric parameters were computed. For the stars with multiple spectra, the error of each parameter is its standard deviation around the mean. For the 9 stars with one single observation, the error is that provided by iSpec. The last two columns indicate whether the star looks very similar (A), similar (B) or different (C) to the Sun in its atmospheric parameters (AP) and Li features at $\sim$670.78 nm.}
  \label{t:ap}
\begin{tabular}{lccclcc}
\hline
HIP     & \teff &  \logg & \mh & N & AP  & Li  \\
\hline
HIP001813        & 5772$\pm$19 &  4.30$\pm$ 0.03 & -0.03$\pm$ 0.01 &  5 & C & A \\
HIP004290        & 5801$\pm$ 9 &  4.58$\pm$ 0.02 & -0.07$\pm$ 0.01 &  5 & C & C \\
HIP007339        & 5630$\pm$12 &  4.36$\pm$ 0.02 &  0.02$\pm$ 0.01 &  1 & C & A \\
HIP007585        & 5779$\pm$10 &  4.41$\pm$ 0.01 &  0.07$\pm$ 0.01 &  5 & B & C \\
HIP007902        & 5631$\pm$11 &  4.36$\pm$ 0.02 & -0.02$\pm$ 0.01 &  1 & C & A \\
HIP007918        & 5889$\pm$ 4 &  4.33$\pm$ 0.01 &  0.05$\pm$ 0.01 &  5 & C & C \\
HIP011253        & 5744$\pm$16 &  4.35$\pm$ 0.02 &  0.01$\pm$ 0.01 &  5 & B & B \\
HIP011728        & 5763$\pm$ 2 &  4.47$\pm$ 0.01 &  0.06$\pm$ 0.01 &  2 & B & A \\
HIP018413        & 5738$\pm$16 &  4.47$\pm$ 0.02 & -0.04$\pm$ 0.01 &  5 & B & A \\
HIP021010        & 5533$\pm$14 &  3.85$\pm$ 0.02 & -0.12$\pm$ 0.01 &  5 & C & C \\
HIP021436        & 5715$\pm$10 &  4.46$\pm$ 0.02 &  0.01$\pm$ 0.01 &  5 & B & A \\
HIP024813        & 5900$\pm$12 &  4.31$\pm$ 0.02 &  0.13$\pm$ 0.01 &  1 & C & C \\
HIP029432        & 5766$\pm$13 &  4.44$\pm$ 0.02 & -0.09$\pm$ 0.01 &  5 & B & B \\
HIP029525        & 5751$\pm$ 8 &  4.60$\pm$ 0.01 & -0.02$\pm$ 0.01 &  5 & C & C \\
HIP031965        & 5782$\pm$15 &  4.28$\pm$ 0.02 &  0.05$\pm$ 0.01 &  5 & C & A \\
HIP035265        & 5857$\pm$ 3 &  4.47$\pm$ 0.01 &  0.03$\pm$ 0.01 &  5 & B & C \\
HIP036874        & 5743$\pm$18 &  4.21$\pm$ 0.03 & -0.07$\pm$ 0.01 &  2 & C & A \\
HIP041484        & 5855$\pm$11 &  4.40$\pm$ 0.02 &  0.03$\pm$ 0.01 &  5 & B & C \\
HIP041844        & 5860$\pm$19 &  4.25$\pm$ 0.02 & -0.03$\pm$ 0.01 &  5 & C & B \\
HIP042575        & 5710$\pm$11 &  4.50$\pm$ 0.02 &  0.05$\pm$ 0.01 &  1 & B & A \\
HIP043557        & 5848$\pm$13 &  4.48$\pm$ 0.02 & -0.04$\pm$ 0.01 &  1 & B & B \\
HIP043726        & 5776$\pm$10 &  4.54$\pm$ 0.02 &  0.13$\pm$ 0.01 &  1 & C & C \\
HIP044089        & 5749$\pm$ 9 &  4.31$\pm$ 0.02 &  0.04$\pm$ 0.01 &  5 & C & A \\
HIP049756        & 5787$\pm$14 &  4.41$\pm$ 0.02 &  0.05$\pm$ 0.01 &  5 & B & B \\
HIP050316        & 5779$\pm$ 9 &  4.07$\pm$ 0.02 &  0.02$\pm$ 0.01 &  5 & C & C \\
HIP050505        & 5661$\pm$ 5 &  4.49$\pm$ 0.01 & -0.14$\pm$ 0.01 &  5 & C & A \\
HIP053721        & 5861$\pm$13 &  4.26$\pm$ 0.01 &  0.01$\pm$ 0.01 &  5 & C & C \\
HIP056832        & 5750$\pm$13 &  4.51$\pm$ 0.02 &  0.06$\pm$ 0.01 &  5 & B & A \\
HIP062527        & 5858$\pm$ 7 &  4.33$\pm$ 0.01 &  0.16$\pm$ 0.01 &  5 & C & B \\
HIP063636        & 5783$\pm$ 9 &  4.58$\pm$ 0.01 & -0.04$\pm$ 0.01 &  5 & C & C \\
HIP064150        & 5726$\pm$20 &  4.32$\pm$ 0.03 &  0.06$\pm$ 0.01 &  5 & C & A \\
HIP065721        & 5501$\pm$ 5 &  3.90$\pm$ 0.01 & -0.09$\pm$ 0.01 &  5 & C & C \\
HIP072043        & 5811$\pm$18 &  4.27$\pm$ 0.02 & -0.02$\pm$ 0.01 &  5 & C & A \\
HIP076114        & 5757$\pm$10 &  4.42$\pm$ 0.01 &  0.00$\pm$ 0.01 &  5 & A & A \\
HIP077052        & 5673$\pm$13 &  4.52$\pm$ 0.01 &  0.07$\pm$ 0.01 &  5 & C & B \\
HIP079672        & 5797$\pm$ 9 &  4.43$\pm$ 0.02 &  0.05$\pm$ 0.01 &  1 & B & B \\
HIP085244        & 5789$\pm$13 &  4.47$\pm$ 0.02 &  0.04$\pm$ 0.01 &  1 & A & B \\
HIP085810        & 5862$\pm$14 &  4.45$\pm$ 0.02 &  0.17$\pm$ 0.01 &  5 & C & C \\
HIP086193        & 5677$\pm$ 9 &  4.26$\pm$ 0.02 &  0.06$\pm$ 0.01 &  5 & C & A \\
HIP088194        & 5759$\pm$ 4 &  4.40$\pm$ 0.01 & -0.05$\pm$ 0.01 &  5 & A & A \\
HIP089474        & 5802$\pm$10 &  4.29$\pm$ 0.01 &  0.02$\pm$ 0.01 &  5 & C & A \\
HIP090355        & 5585$\pm$ 3 &  4.45$\pm$ 0.01 & -0.14$\pm$ 0.01 &  3 & C & A \\
HIP094981        & 5693$\pm$ 7 &  4.52$\pm$ 0.01 &  0.06$\pm$ 0.01 &  5 & B & C \\
HIP096402        & 5612$\pm$ 2 &  4.17$\pm$ 0.01 & -0.06$\pm$ 0.01 &  2 & C & A \\
HIP096895        & 5827$\pm$12 &  4.36$\pm$ 0.02 &  0.15$\pm$ 0.01 &  1 & C & A \\
HIP096901        & 5735$\pm$ 9 &  4.25$\pm$ 0.04 &  0.06$\pm$ 0.01 &  2 & C & B \\
HIP096948        & 5761$\pm$14 &  4.39$\pm$ 0.01 &  0.11$\pm$ 0.01 &  3 & C & B \\
HIP097336        & 5812$\pm$13 &  4.36$\pm$ 0.01 &  0.13$\pm$ 0.02 &  5 & C & A \\
HIP097420        & 5798$\pm$11 &  4.50$\pm$ 0.02 &  0.03$\pm$ 0.01 &  5 & B & C \\
HIP097767        & 5670$\pm$ 5 &  4.03$\pm$ 0.01 & -0.16$\pm$ 0.01 &  5 & C & A \\
HIP100963        & 5821$\pm$ 6 &  4.49$\pm$ 0.01 &  0.01$\pm$ 0.01 &  5 & B & C \\
HIP102040        & 5840$\pm$ 4 &  4.45$\pm$ 0.01 & -0.07$\pm$ 0.01 &  5 & B & C \\
HIP116613        & 5832$\pm$10 &  4.57$\pm$ 0.01 &  0.13$\pm$ 0.01 &  5 & C & C \\
HIP118162        & 5753$\pm$ 4 &  4.44$\pm$ 0.01 &  0.08$\pm$ 0.01 &  2 & B & A \\
TYC2583-01846-2  & 5890$\pm$ 6 &  4.53$\pm$ 0.01 & -0.03$\pm$ 0.01 &  5 & C & C \\
TYC2694-00364-1  & 5842$\pm$ 1 &  4.44$\pm$ 0.01 &  0.03$\pm$ 0.01 &  5 & B & C \\
\hline
\end{tabular}
\end{table*}

\section{Li content }
\label{s:li}
\begin{figure*}[t!]
\begin{center}

\includegraphics [width=6 cm] {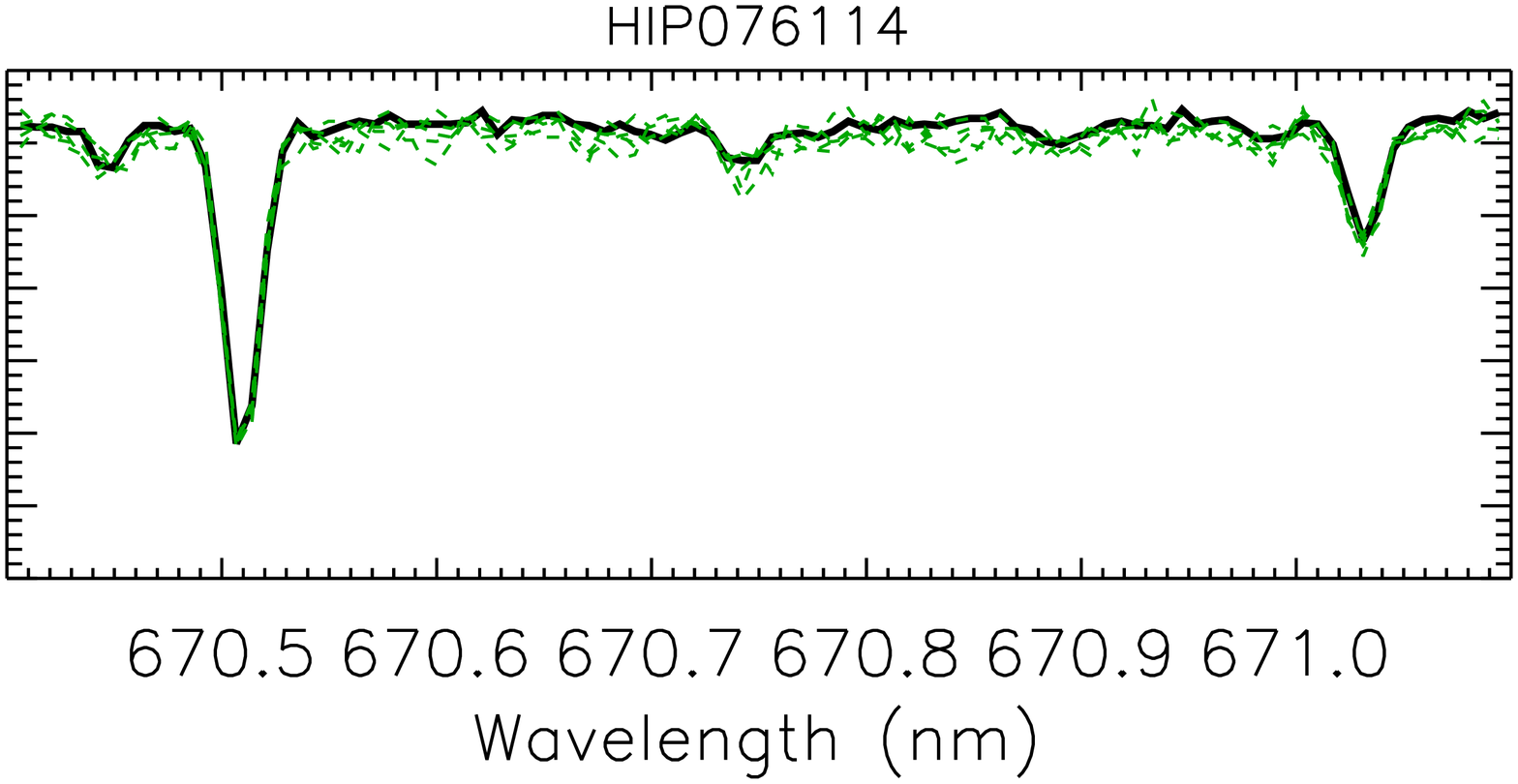} \includegraphics [width=6 cm] {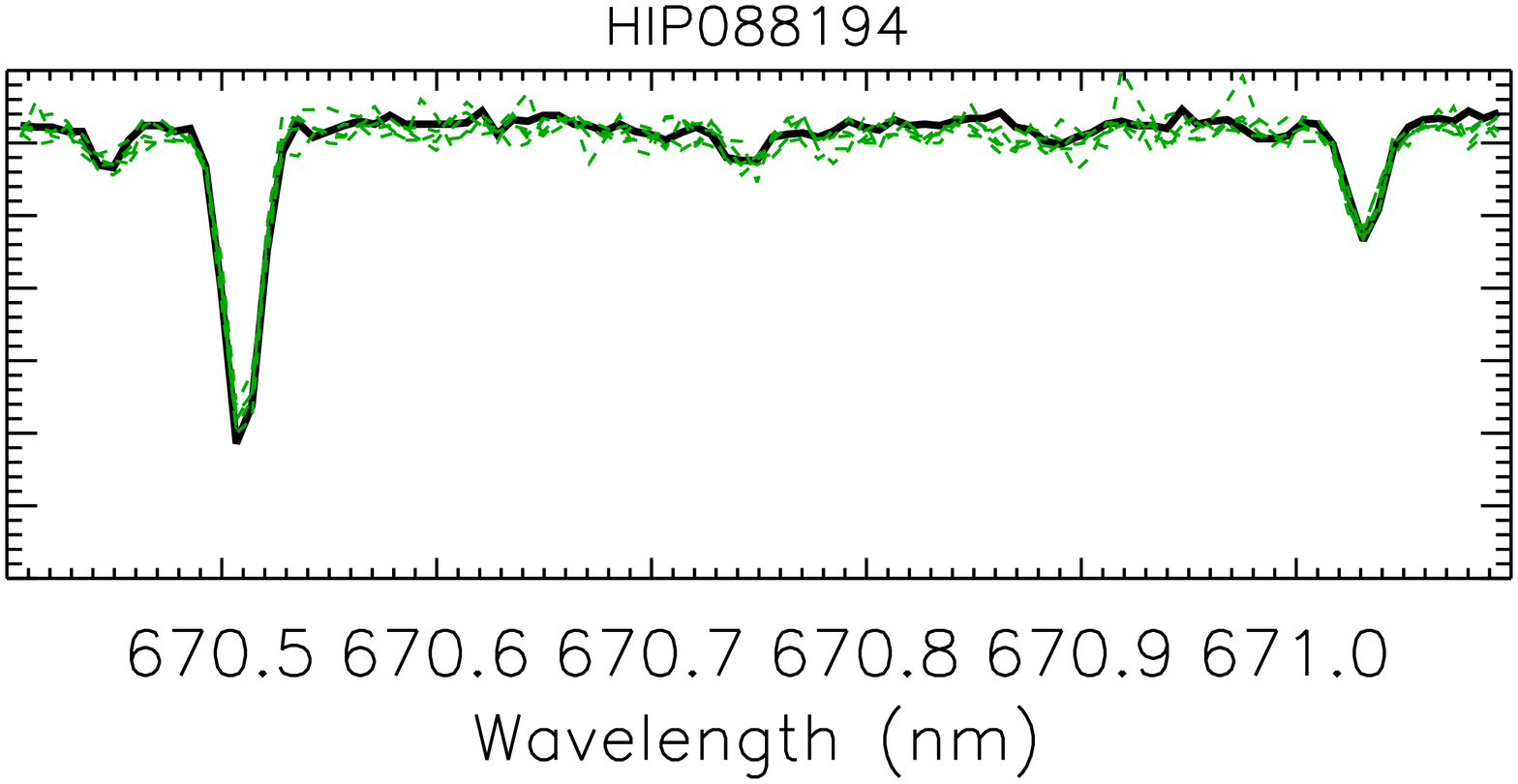}
\includegraphics [width=6 cm] {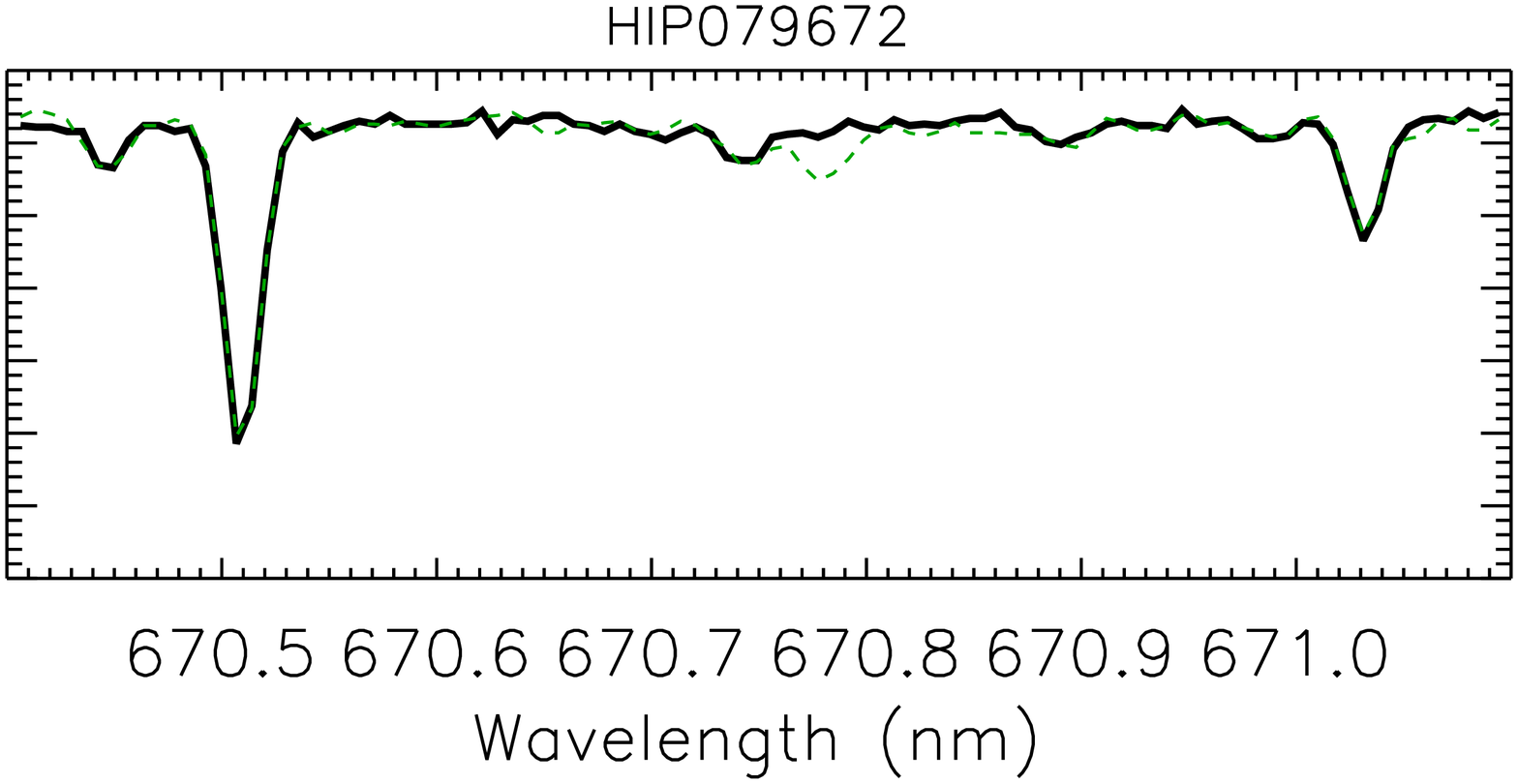}\includegraphics [width=6 cm] {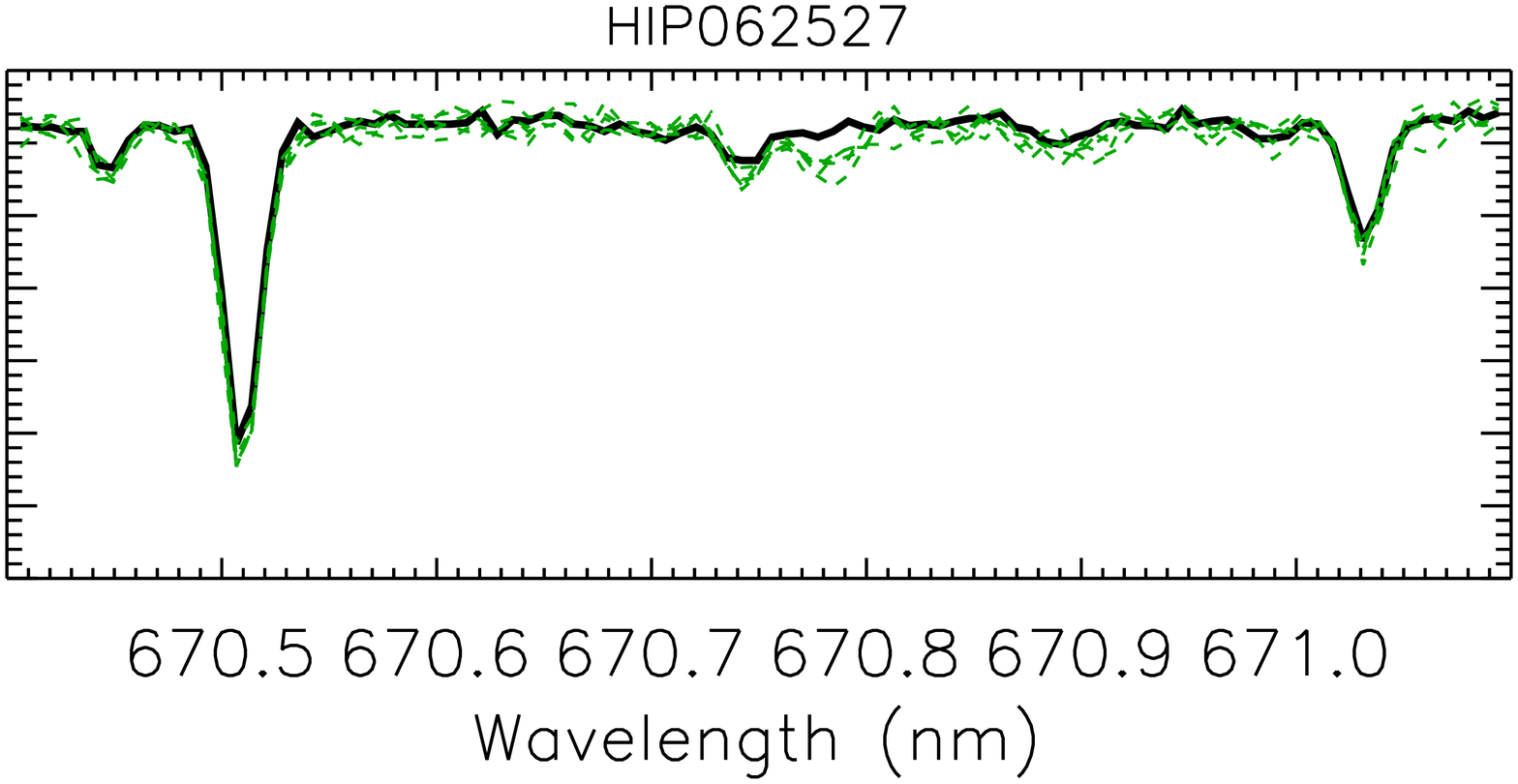}
\includegraphics [width=6 cm] {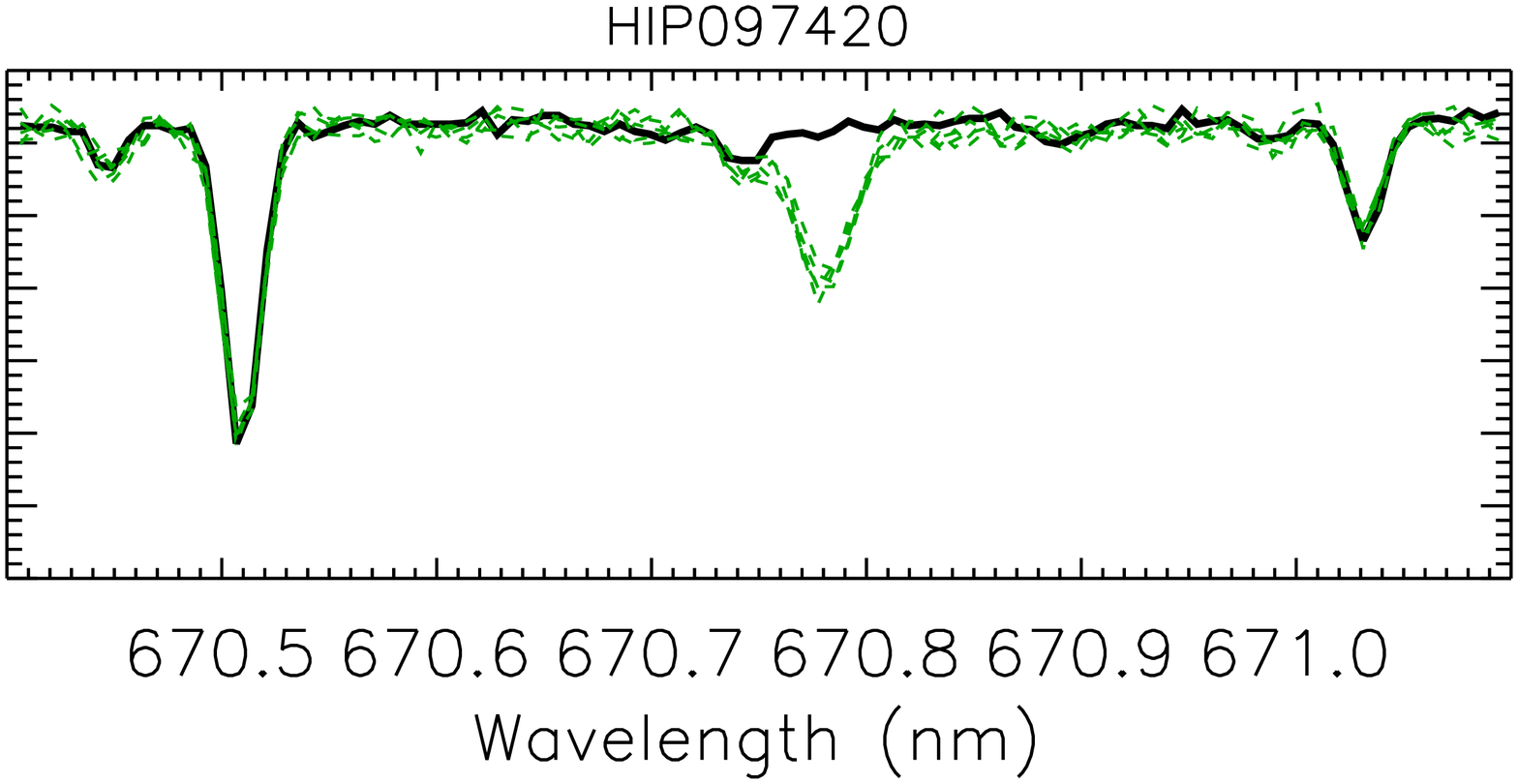}\includegraphics [width=6 cm] {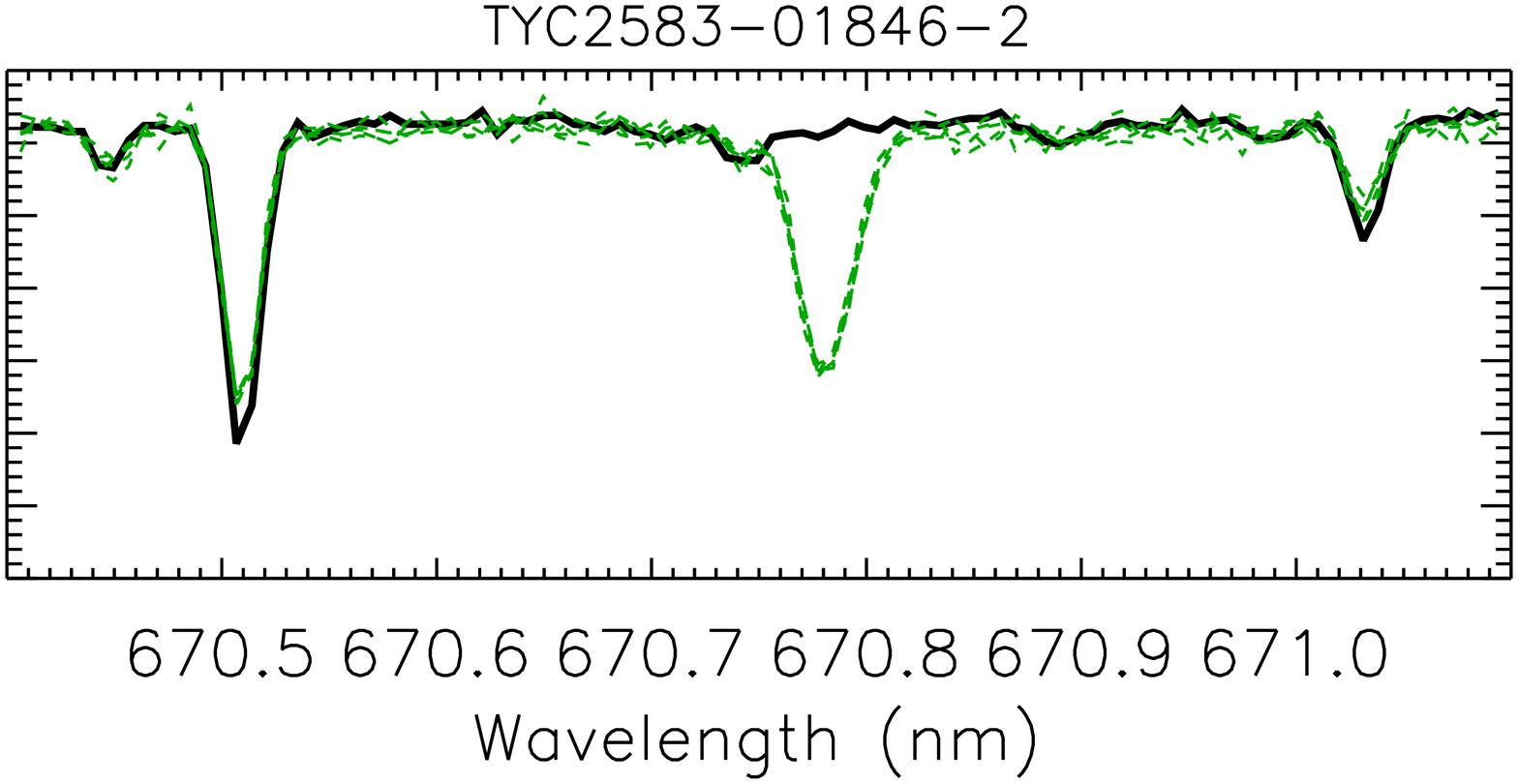}

\caption{Comparison of the spectral range centered on the 670.78 nm Li feature in one of the Sun spectra  (black thick line) and a target (green dashed line), with up to 5 spectra available for some stars. The upper panels show two stars classified A (same Li deficiency as the Sun), the middle panels show two stars classified B (slightly more abundant in Li than the Sun), the bottom panels show two stars classified C (pronounced Li feature).   }
\label{f:li}
\end{center}
\end{figure*}

It is known that the lithium abundances of solar type stars show a large dispersion \citep[see for instance the comprehensive study by][]{2007A&A...468..663T}. This is subject to various interpretations.  The Sun is depleted in Li and it is expected that solar twins show the same deficiency. We have classified our targets stars into three categories after a visual inspection of the wavelength range inluding the Li lines at $\sim$670.78 nm in comparison to one of the \elodie\ solar spectra. We classified into A in Table \ref{t:ap} the stars which show  a similar Li deficiency as the Sun, into B those with a slightly higher Li abundance and into C those which exhibit a pronounced Li feature. There are 24, 11 and 21 stars in categories A, B and C respectively. Figure \ref{f:li} shows two examples of stars in each category. It is worth to note that HIP076114 and  HIP088194 are classified A for their similarity to the Sun in both atmospheric parameters and Lithium content. 

Stellar rotation and activity are strongly correlated with the surface Li content of solar type stars \citep{2007A&A...468..663T}. The stars examined here are supposedly slow rotators due to our initial constraint FWHM $\le$ 12 \kms. It is thus interesting to find so diverse strengths of their Li  feature.

\section{Abundances}
\label{s:ab}
    The chemical abundances  were derived with iSpec and the line list of the Gaia ESO survey, this time including also lines labeled to be less reliable, due to uncertain atomic data or possible blends. However the 18 solar spectra allowed us to keep only the lines which gave consistent abundances from one spectrum to another. Moreover since we are dealing with solar twins, a differential analysis with respect to solar spectra was possible, reducing the uncertainty in relative abundances due to bad atomic data. In that way we were able to obtain a high precision, which is so important to identify solar twins \citep{2012A&A...543A..29M}. 

\begin{figure}[h!]
\begin{center}
 \includegraphics [width=6 cm] {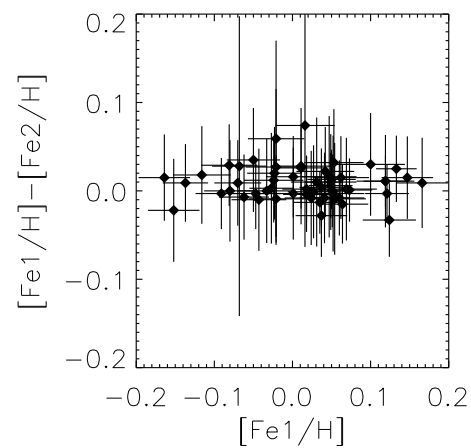} 
\caption{Difference of iron abundances obtained with neutral and ionised lines for the 56 twin candidates. The mean difference is 0.008 dex with a standard deviation of 0.019 dex. }
\label{f:feh}
\end{center}
\end{figure}

A total of 1865 lines were measurable in the \elodie\ spectra out of which were kept only those measured in at least 17 out of 18 solar spectra with a weighted standard deviation lower than 0.02. Then we selected the chemical elements for which at least 3 such lines were measured. This left us with 189 Fe lines, 23 Ni lines, 15 Si and Ca lines, 8 Cr lines, 5 Mn and Ti lines, 4 Na lines, all neutral, plus 3 Fe$\rm{II}$ lines.   Table \ref{t:lines} lists these 267 lines which were found to be suitable for spectral synthesis of solar type stars. For each of these lines, we took as reference abundance the weighted mean obtained on the 17 or 18 solar spectra. The line by line abundances of the 225 spectra were then measured differentially for this set. As the total error for a line abundance relative to the Sun, we quadratically summed the weighted standard deviation obtained for the solar spectra and the rms of the fit obtained for the target spectrum. Then the weighted average and standard deviation were computed for each element and each star, as presented in Tables \ref{t:fena}, \ref{t:sicati} and \ref{t:crmnni}. With up to 5 spectra available for a large fraction of the targets, the determination of [Fe$\rm{I}$/H] was often based on the  measurement of 945 individual lines. In total we obtained 59\,707 individual line measurements with errors between 0.01 and 0.111 and a median error of  0.02. As a verification, we checked the agreement of the [Fe$\rm{I}$/H] and [Fe$\rm{II}$/H] determinations (Fig. \ref{f:feh}). The mean difference is 0.008, demonstrating the excellent agreement of the 2 determinations of iron abundances,  with a standard deviation of 0.019, agreeing perfectly with the median error of 0.02 of all the individual line measurements. 

Fig. \ref{f:feh} also shows that there is a concentration of stars slightly more metal-rich than the Sun, a possible bias of the content of the \elodie\ archive in favour of stars followed-up for the search of extra-solar planets.

HIP079672 (18 Sco) has been studied at very high resolution (R $\sim$ 110\,000) and S/N (800 - 1000) by \cite{2014ApJ...791...14M} who obtained abundances  differentially to the Sun with an unprecedented level of precision. We compare their determinations to ours in Table \ref{t:18sco} : they agree within 0.03 for Ti and Cr, within 0.008 for Na and Ca and within 0.004 for Fe, Si, Mn and Ni. Thus the agreement is largely better than expected from our quoted errors. For comparison we also show the abundances from \cite{2015arXiv150700027J}  who made an extensive study of the Gaia benchmark stars. They are also in excellent agreement.

\begin{table}[h]
  \centering 
  \caption{Comparison of abundances determined in this work for HIP079672 (18 Sco) to those obtained by \cite{2014ApJ...791...14M} and by \cite{2015arXiv150700027J}.}
  \label{t:18sco}
\begin{tabular}{lccc}
\hline
\hline
[X/H]     & This work &  \cite{2014ApJ...791...14M} & \cite{2015arXiv150700027J}\\
\hline
Fe & 0.051$\pm$0.026 & 0.054$\pm$0.005 &  0.03$\pm$ 0.01 (a) \\
Na  &0.017$\pm$0.045 & 0.025$\pm$0.004 & \\
Si  &0.043$\pm$0.014 & 0.047$\pm$0.003 & 0.048$\pm$  0.018\\
Ca &0.065$\pm$0.016 & 0.057$\pm$0.006 & 0.058$\pm$  0.036\\
Ti  &0.022$\pm$0.045 & 0.051$\pm$0.007 & 0.046$\pm$ 0.026\\
Cr &0.034$\pm$0.026 & 0.056$\pm$0.006 & 0.049$\pm$ 0.023\\
Mn &0.037$\pm$0.012 & 0.041$\pm$0.006 & 0.040$\pm$ 0.018\\
Ni &0.038$\pm$0.027 & 0.041$\pm$0.004 &0.039$\pm$ 0.017 \\
\hline
\end{tabular}
\tablefoot{(a) NLTE determination from \cite{2014A&A...564A.133J}}
\end{table}

Seven solar analogs have all the tested elements agreeing with the solar abundances within 0.05 : HIP021436, HIP035265, HIP076114, HIP085244, HIP089474, HIP100963, TYC2694-00364-1.

\begin{figure*}[h!]
\begin{center}
 \includegraphics [width=18cm] {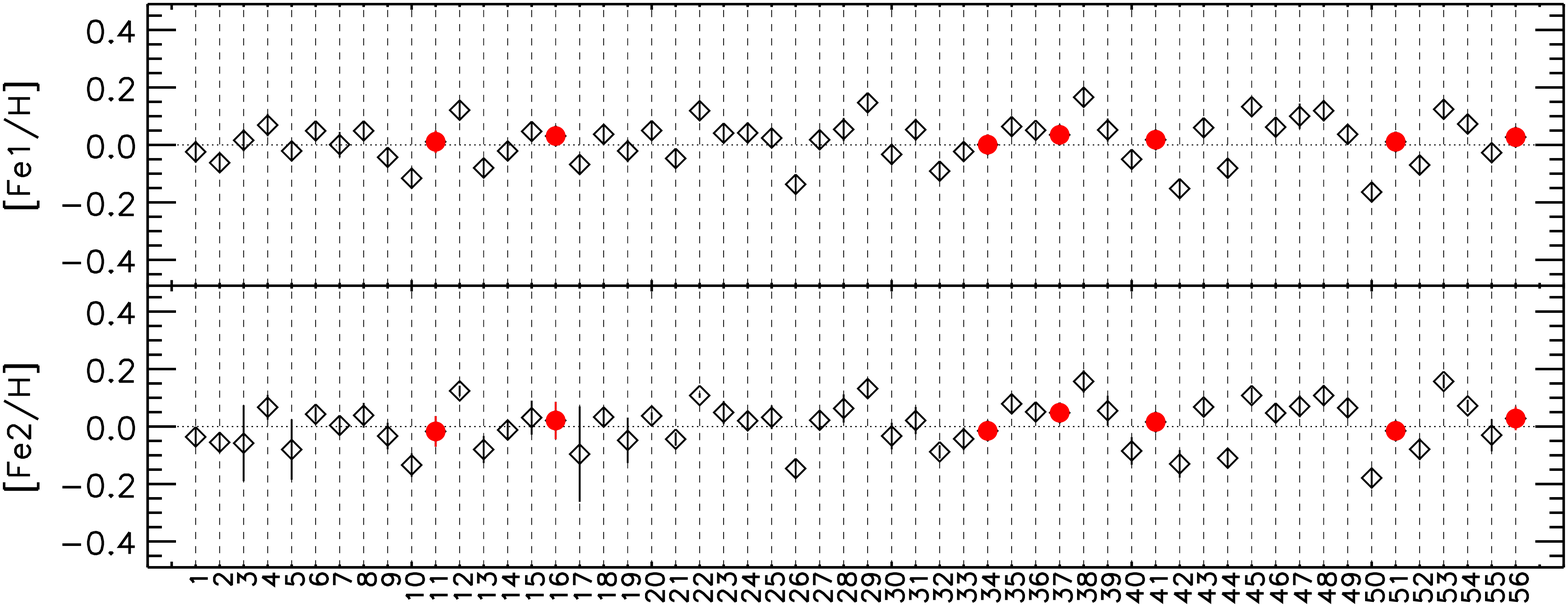} 
\caption{Iron abundance of the 56 solar twin candidates, obtained from neutral and ionized lines. Red dots represent the 7 stars found to have the same abundances than the Sun, within 0.05 dex, for all the tested elements.}
\label{f:fe}
\end{center}
\end{figure*}

\begin{figure*}[h!]
\begin{center}
\includegraphics [width=18 cm] {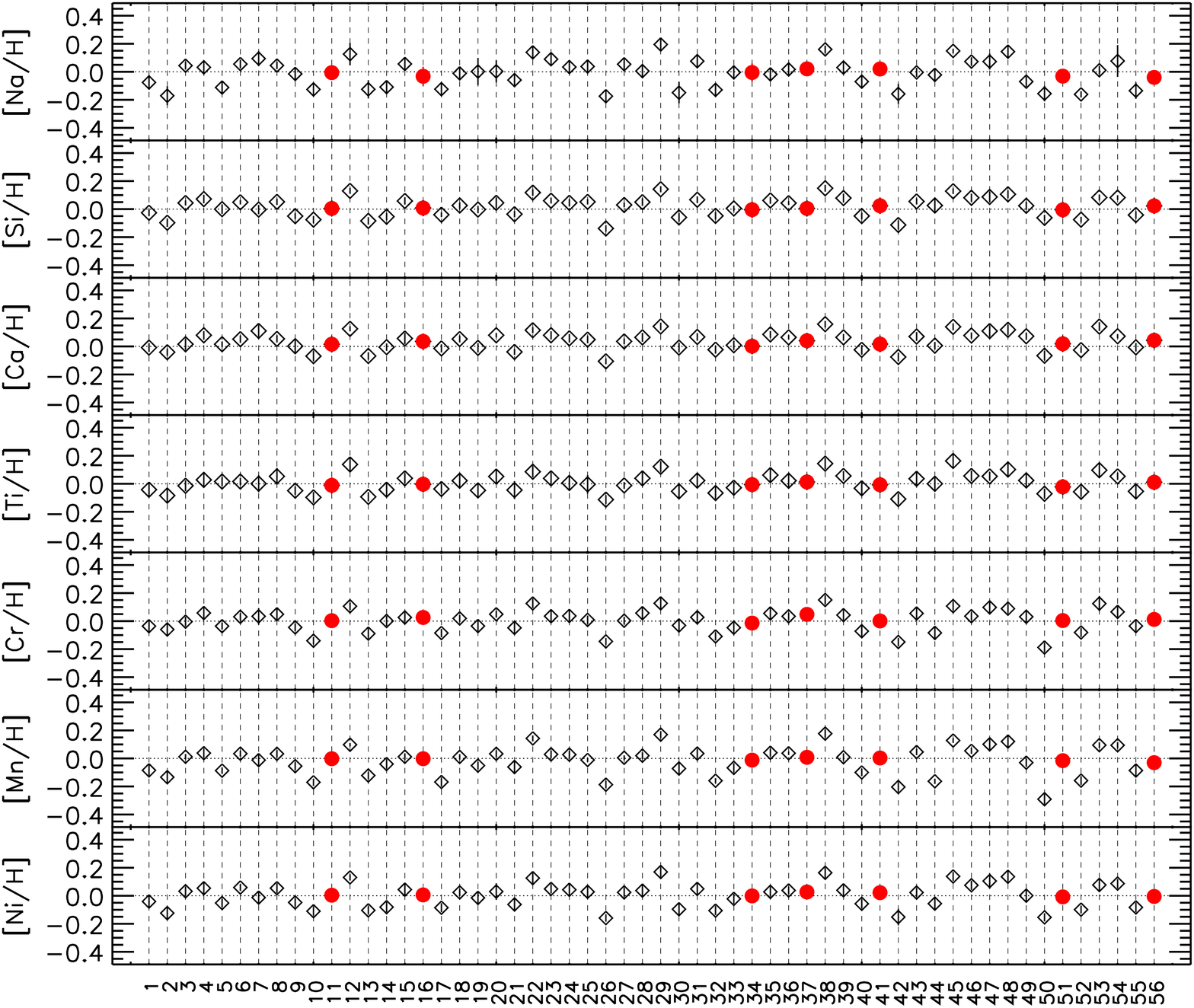} 
\caption{Like Fig. \ref{f:fe} for the other elements.}
\label{f:el}
\end{center}
\end{figure*}

Figure \ref{f:fe} shows the iron abundances of the 56 targets from neutral and ionized lines and Fig. \ref{f:el} displays the abundances of the other elements. The stars are numbered from 1 to 56 for  the sake of clarity. The abundances of the tested elements are quite well centered on the solar value, with a small spread. This suggests that the solar abundance pattern is not unusual as suggested by some studies \citep{2015A&A...579A..52N}. This point needs however further confirmation with other elements which are not part of this study because of our strict selection of lines very well measured in the \elodie\ solar spectra, which has  guaranteed a high level of precision.

\begin{table*}[h]
  \centering 
  \caption{Detailed abundances of the 56 targets }
  \label{t:fena}
\begin{tabular}{clrrrrrrrrr}
\hline
\hline
num & HIP/TYC  &  \multicolumn{3}{c}{[Fe$\rm{I}$/H]} &   \multicolumn{3}{c}{[Fe$\rm{II}$/H]} &      \multicolumn{3}{c}{[Na/H]} \\
 & & \small{Mean} & $\sigma$ & N & \small{Mean} & $\sigma$ & N &\small{Mean} & $\sigma$ & N \\
 \hline
  1 & HIP001813       & -0.024 & 0.038 &  928 & -0.036 & 0.029 &   15 & -0.076 & 0.052 &   16 \\
  2 & HIP004290       & -0.062 & 0.032 &  938 & -0.055 & 0.036 &   15 & -0.171 & 0.067 &   17 \\
  3 & HIP007339       &  0.016 & 0.038 &  185 & -0.058 & 0.132 &    3 &  0.044 & 0.028 &    3 \\
  4 & HIP007585       &  0.069 & 0.039 &  926 &  0.067 & 0.043 &   15 &  0.032 & 0.044 &   15 \\
  5 & HIP007902       & -0.021 & 0.036 &  187 & -0.080 & 0.105 &    3 & -0.112 & 0.041 &    4 \\
  6 & HIP007918       &  0.049 & 0.028 &  944 &  0.043 & 0.024 &   15 &  0.055 & 0.031 &   20 \\
  7 & HIP011253       &  0.001 & 0.044 &  940 &  0.004 & 0.028 &   15 &  0.094 & 0.054 &   19 \\
  8 & HIP011728       &  0.049 & 0.033 &  376 &  0.039 & 0.043 &    6 &  0.045 & 0.026 &    8 \\
  9 & HIP018413       & -0.043 & 0.035 &  939 & -0.033 & 0.046 &   15 & -0.015 & 0.044 &   18 \\
 10 & HIP021010       & -0.116 & 0.037 &  940 & -0.134 & 0.041 &   15 & -0.126 & 0.042 &   20 \\
 11 & HIP021436       &  0.011 & 0.039 &  939 & -0.017 & 0.053 &   15 & -0.006 & 0.047 &   18 \\
 12 & HIP024813       &  0.121 & 0.028 &  187 &  0.124 & 0.018 &    3 &  0.126 & 0.080 &    4 \\
 13 & HIP029432       & -0.080 & 0.033 &  933 & -0.080 & 0.047 &   15 & -0.125 & 0.065 &   16 \\
 14 & HIP029525       & -0.021 & 0.035 &  939 & -0.012 & 0.035 &   15 & -0.109 & 0.040 &   19 \\
 15 & HIP031965       &  0.047 & 0.035 &  935 &  0.031 & 0.059 &   15 &  0.056 & 0.043 &   20 \\
 16 & HIP035265       &  0.031 & 0.033 &  934 &  0.021 & 0.065 &   15 & -0.032 & 0.068 &   20 \\
 17 & HIP036874       & -0.068 & 0.035 &  366 & -0.096 & 0.166 &    6 & -0.124 & 0.039 &    8 \\
 18 & HIP041484       &  0.037 & 0.028 &  940 &  0.033 & 0.034 &   15 & -0.011 & 0.044 &   18 \\
 19 & HIP041844       & -0.021 & 0.039 &  938 & -0.048 & 0.079 &   15 &  0.002 & 0.097 &   20 \\
 20 & HIP042575       &  0.050 & 0.031 &  186 &  0.037 & 0.013 &    3 &  0.003 & 0.078 &    4 \\
 21 & HIP043557       & -0.047 & 0.033 &  189 & -0.044 & 0.023 &    3 & -0.059 & 0.044 &    4 \\
 22 & HIP043726       &  0.119 & 0.028 &  188 &  0.108 & 0.008 &    3 &  0.139 & 0.040 &    4 \\
 23 & HIP044089       &  0.041 & 0.032 &  943 &  0.049 & 0.040 &   15 &  0.090 & 0.039 &   19 \\
 24 & HIP049756       &  0.042 & 0.037 &  938 &  0.020 & 0.034 &   15 &  0.034 & 0.047 &   18 \\
 25 & HIP050316       &  0.024 & 0.034 &  934 &  0.032 & 0.041 &   15 &  0.038 & 0.045 &   20 \\
 26 & HIP050505       & -0.137 & 0.029 &  945 & -0.146 & 0.033 &   15 & -0.174 & 0.039 &   21 \\
 27 & HIP053721       &  0.018 & 0.030 &  940 &  0.022 & 0.030 &   15 &  0.054 & 0.037 &   20 \\
 28 & HIP056832       &  0.054 & 0.040 &  938 &  0.063 & 0.049 &   15 &  0.005 & 0.041 &   20 \\
 29 & HIP062527       &  0.147 & 0.033 &  940 &  0.132 & 0.033 &   15 &  0.195 & 0.038 &   19 \\
 30 & HIP063636       & -0.033 & 0.037 &  941 & -0.033 & 0.046 &   15 & -0.149 & 0.076 &   20 \\
 31 & HIP064150       &  0.053 & 0.038 &  939 &  0.021 & 0.047 &   15 &  0.076 & 0.050 &   20 \\
 32 & HIP065721       & -0.091 & 0.033 &  939 & -0.088 & 0.023 &   15 & -0.129 & 0.044 &   19 \\
 33 & HIP072043       & -0.023 & 0.035 &  942 & -0.043 & 0.039 &   15 & -0.004 & 0.047 &   19 \\
 34 & HIP076114       &  0.001 & 0.035 &  928 & -0.015 & 0.030 &   15 & -0.005 & 0.063 &   18 \\
 35 & HIP077052       &  0.064 & 0.033 &  942 &  0.079 & 0.030 &   15 & -0.018 & 0.046 &   20 \\
 36 & HIP079672       &  0.051 & 0.026 &  188 &  0.051 & 0.032 &    3 &  0.017 & 0.045 &    4 \\
 37 & HIP085244       &  0.035 & 0.034 &  188 &  0.048 & 0.010 &    3 &  0.021 & 0.031 &    4 \\
 38 & HIP085810       &  0.166 & 0.033 &  923 &  0.157 & 0.039 &   15 &  0.160 & 0.027 &   18 \\
 39 & HIP086193       &  0.052 & 0.040 &  923 &  0.055 & 0.052 &   15 &  0.030 & 0.045 &   18 \\
 40 & HIP088194       & -0.050 & 0.034 &  945 & -0.085 & 0.048 &   15 & -0.070 & 0.040 &   20 \\
 41 & HIP089474       &  0.018 & 0.031 &  925 &  0.016 & 0.022 &   15 &  0.020 & 0.029 &   18 \\
 42 & HIP090355       & -0.152 & 0.033 &  566 & -0.130 & 0.048 &    9 & -0.158 & 0.071 &   12 \\
 43 & HIP094981       &  0.060 & 0.032 &  945 &  0.068 & 0.028 &   15 & -0.004 & 0.032 &   20 \\
 44 & HIP096402       & -0.081 & 0.036 &  377 & -0.110 & 0.029 &    6 & -0.022 & 0.042 &    8 \\
 45 & HIP096895       &  0.133 & 0.026 &  184 &  0.108 & 0.027 &    3 &  0.149 & 0.016 &    3 \\
 46 & HIP096901       &  0.062 & 0.036 &  374 &  0.047 & 0.030 &    6 &  0.073 & 0.031 &    6 \\
 47 & HIP096948       &  0.100 & 0.044 &  555 &  0.070 & 0.038 &    9 &  0.073 & 0.057 &   11 \\
 48 & HIP097336       &  0.119 & 0.037 &  941 &  0.108 & 0.037 &   15 &  0.144 & 0.041 &   18 \\
 49 & HIP097420       &  0.037 & 0.033 &  936 &  0.065 & 0.033 &   15 & -0.070 & 0.036 &   17 \\
 50 & HIP097767       & -0.164 & 0.033 &  940 & -0.179 & 0.036 &   15 & -0.156 & 0.055 &   19 \\
 51 & HIP100963       &  0.011 & 0.034 &  946 & -0.015 & 0.031 &   15 & -0.031 & 0.034 &   20 \\
 52 & HIP102040       & -0.070 & 0.030 &  932 & -0.079 & 0.038 &   15 & -0.161 & 0.022 &   17 \\
 53 & HIP116613       &  0.124 & 0.034 &  944 &  0.157 & 0.024 &   15 &  0.011 & 0.030 &   19 \\
 54 & HIP118162       &  0.073 & 0.029 &  378 &  0.072 & 0.021 &    6 &  0.077 & 0.113 &    8 \\
 55 & TYC2583-01846-2 & -0.027 & 0.028 &  943 & -0.030 & 0.056 &   15 & -0.135 & 0.058 &   19 \\
 56 & TYC2694-00364-1 &  0.027 & 0.033 &  933 &  0.028 & 0.041 &   15 & -0.040 & 0.022 &   15 \\
\hline
\end{tabular}
\end{table*}

  \begin{table*}[h]
  \centering 
  \caption{Detailed abundances of the 56 targets }
  \label{t:sicati}
\begin{tabular}{clrrrrrrrrr}
\hline
\hline
num & HIP/TYC  &  \multicolumn{3}{c}{[Si/H]} &   \multicolumn{3}{c}{[Ca/H]} &     \multicolumn{3}{c}{[Ti/H]} \\
 & & \small{Mean} & $\sigma$ & N & \small{Mean} & $\sigma$ & N &\small{Mean} & $\sigma$ & N   \\
 \hline
  1 & HIP001813       & -0.026 & 0.024 &   75 & -0.008 & 0.031 &   75 & -0.042 & 0.049 &   25 \\
  2 & HIP004290       & -0.097 & 0.030 &   75 & -0.041 & 0.025 &   75 & -0.084 & 0.054 &   25 \\
  3 & HIP007339       &  0.044 & 0.018 &   15 &  0.016 & 0.038 &   15 & -0.014 & 0.039 &    5 \\
  4 & HIP007585       &  0.072 & 0.030 &   75 &  0.080 & 0.029 &   75 &  0.028 & 0.046 &   25 \\
  5 & HIP007902       &  0.000 & 0.034 &   15 &  0.016 & 0.020 &   15 &  0.016 & 0.032 &    5 \\
  6 & HIP007918       &  0.049 & 0.016 &   75 &  0.052 & 0.026 &   75 &  0.016 & 0.026 &   25 \\
  7 & HIP011253       & -0.003 & 0.033 &   75 &  0.111 & 0.041 &   75 &  0.001 & 0.061 &   25 \\
  8 & HIP011728       &  0.052 & 0.022 &   30 &  0.054 & 0.025 &   30 &  0.053 & 0.060 &   10 \\
  9 & HIP018413       & -0.050 & 0.021 &   75 &  0.002 & 0.043 &   75 & -0.049 & 0.029 &   25 \\
 10 & HIP021010       & -0.075 & 0.026 &   75 & -0.069 & 0.038 &   75 & -0.097 & 0.039 &   25 \\
 11 & HIP021436       &  0.005 & 0.030 &   75 &  0.016 & 0.031 &   75 & -0.011 & 0.043 &   25 \\
 12 & HIP024813       &  0.130 & 0.021 &   15 &  0.126 & 0.018 &   15 &  0.138 & 0.045 &    5 \\
 13 & HIP029432       & -0.084 & 0.026 &   74 & -0.068 & 0.027 &   75 & -0.093 & 0.033 &   25 \\
 14 & HIP029525       & -0.054 & 0.026 &   75 & -0.005 & 0.026 &   75 & -0.042 & 0.051 &   25 \\
 15 & HIP031965       &  0.058 & 0.026 &   75 &  0.057 & 0.040 &   75 &  0.039 & 0.037 &   25 \\
 16 & HIP035265       &  0.008 & 0.025 &   75 &  0.036 & 0.023 &   75 & -0.003 & 0.029 &   25 \\
 17 & HIP036874       & -0.040 & 0.017 &   28 & -0.014 & 0.029 &   28 & -0.037 & 0.033 &   10 \\
 18 & HIP041484       &  0.026 & 0.022 &   75 &  0.053 & 0.024 &   75 &  0.022 & 0.044 &   25 \\
 19 & HIP041844       & -0.003 & 0.026 &   75 & -0.010 & 0.033 &   75 & -0.047 & 0.043 &   25 \\
 20 & HIP042575       &  0.045 & 0.038 &   15 &  0.080 & 0.016 &   15 &  0.052 & 0.047 &    5 \\
 21 & HIP043557       & -0.036 & 0.022 &   15 & -0.039 & 0.021 &   15 & -0.045 & 0.064 &    5 \\
 22 & HIP043726       &  0.118 & 0.024 &   15 &  0.116 & 0.022 &   15 &  0.086 & 0.040 &    5 \\
 23 & HIP044089       &  0.060 & 0.020 &   75 &  0.080 & 0.028 &   75 &  0.039 & 0.040 &   26 \\
 24 & HIP049756       &  0.045 & 0.026 &   75 &  0.058 & 0.030 &   75 &  0.006 & 0.047 &   25 \\
 25 & HIP050316       &  0.052 & 0.019 &   75 &  0.049 & 0.026 &   75 & -0.004 & 0.066 &   25 \\
 26 & HIP050505       & -0.138 & 0.023 &   75 & -0.106 & 0.021 &   75 & -0.113 & 0.031 &   25 \\
 27 & HIP053721       &  0.030 & 0.017 &   75 &  0.036 & 0.021 &   75 & -0.012 & 0.024 &   25 \\
 28 & HIP056832       &  0.050 & 0.027 &   75 &  0.065 & 0.031 &   75 &  0.039 & 0.047 &   25 \\
 29 & HIP062527       &  0.142 & 0.022 &   75 &  0.143 & 0.027 &   75 &  0.122 & 0.037 &   25 \\
 30 & HIP063636       & -0.059 & 0.030 &   75 & -0.008 & 0.035 &   75 & -0.054 & 0.057 &   25 \\
 31 & HIP064150       &  0.067 & 0.024 &   75 &  0.068 & 0.027 &   75 &  0.023 & 0.044 &   25 \\
 32 & HIP065721       & -0.048 & 0.026 &   75 & -0.023 & 0.028 &   75 & -0.066 & 0.043 &   25 \\
 33 & HIP072043       &  0.005 & 0.020 &   75 &  0.009 & 0.032 &   75 & -0.028 & 0.042 &   25 \\
 34 & HIP076114       & -0.004 & 0.023 &   74 &  0.002 & 0.029 &   75 & -0.006 & 0.055 &   25 \\
 35 & HIP077052       &  0.062 & 0.024 &   75 &  0.086 & 0.024 &   75 &  0.062 & 0.048 &   25 \\
 36 & HIP079672       &  0.043 & 0.014 &   15 &  0.065 & 0.016 &   15 &  0.022 & 0.045 &    5 \\
 37 & HIP085244       &  0.005 & 0.021 &   15 &  0.041 & 0.028 &   15 &  0.013 & 0.052 &    5 \\
 38 & HIP085810       &  0.149 & 0.024 &   74 &  0.159 & 0.023 &   75 &  0.144 & 0.040 &   25 \\
 39 & HIP086193       &  0.079 & 0.032 &   72 &  0.065 & 0.024 &   74 &  0.056 & 0.034 &   25 \\
 40 & HIP088194       & -0.049 & 0.023 &   75 & -0.024 & 0.034 &   75 & -0.032 & 0.044 &   25 \\
 41 & HIP089474       &  0.024 & 0.019 &   74 &  0.017 & 0.036 &   75 & -0.007 & 0.022 &   25 \\
 42 & HIP090355       & -0.113 & 0.033 &   45 & -0.075 & 0.031 &   45 & -0.109 & 0.039 &   15 \\
 43 & HIP094981       &  0.055 & 0.022 &   75 &  0.071 & 0.031 &   75 &  0.036 & 0.046 &   25 \\
 44 & HIP096402       &  0.026 & 0.048 &   30 &  0.006 & 0.026 &   30 &  0.000 & 0.044 &   10 \\
 45 & HIP096895       &  0.128 & 0.020 &   15 &  0.140 & 0.025 &   15 &  0.163 & 0.046 &    5 \\
 46 & HIP096901       &  0.081 & 0.029 &   30 &  0.078 & 0.022 &   30 &  0.056 & 0.038 &   10 \\
 47 & HIP096948       &  0.087 & 0.038 &   43 &  0.110 & 0.039 &   45 &  0.053 & 0.032 &   15 \\
 48 & HIP097336       &  0.106 & 0.027 &   75 &  0.119 & 0.037 &   75 &  0.102 & 0.055 &   25 \\
 49 & HIP097420       &  0.024 & 0.028 &   75 &  0.073 & 0.026 &   75 &  0.024 & 0.054 &   25 \\
 50 & HIP097767       & -0.062 & 0.021 &   75 & -0.066 & 0.034 &   75 & -0.071 & 0.036 &   25 \\
 51 & HIP100963       & -0.005 & 0.021 &   75 &  0.019 & 0.026 &   75 & -0.022 & 0.039 &   25 \\
 52 & HIP102040       & -0.075 & 0.020 &   75 & -0.026 & 0.020 &   75 & -0.057 & 0.036 &   25 \\
 53 & HIP116613       &  0.083 & 0.021 &   75 &  0.142 & 0.031 &   75 &  0.097 & 0.043 &   25 \\
 54 & HIP118162       &  0.082 & 0.017 &   30 &  0.073 & 0.015 &   30 &  0.053 & 0.026 &   10 \\
 55 & TYC2583-01846-2 & -0.042 & 0.026 &   76 & -0.006 & 0.030 &   75 & -0.054 & 0.046 &   25 \\
 56 & TYC2694-00364-1 &  0.023 & 0.030 &   75 &  0.044 & 0.031 &   75 &  0.011 & 0.036 &   25 \\
\hline
\end{tabular}
\end{table*}
   
\begin{table*}[h]
  \centering 
  \caption{Detailed abundances of the 56 targets }
  \label{t:crmnni}
\begin{tabular}{clrrrrrrrrr}
\hline
\hline
num & HIP/TYC  &  \multicolumn{3}{c}{[Cr/H]} &   \multicolumn{3}{c}{[Mn/H]} &     \multicolumn{3}{c}{[Ni/H]} \\
 & & \small{Mean} & $\sigma$ & N & \small{Mean} & $\sigma$ & N &\small{Mean} & $\sigma$ & N   \\
 \hline
   1 & HIP001813       & -0.036 & 0.037 &   40 & -0.085 & 0.037 &   25 & -0.041 & 0.038 &  115 \\
  2 & HIP004290       & -0.060 & 0.040 &   40 & -0.134 & 0.038 &   25 & -0.124 & 0.040 &  115 \\
  3 & HIP007339       & -0.004 & 0.035 &    8 &  0.012 & 0.019 &    5 &  0.032 & 0.032 &   23 \\
  4 & HIP007585       &  0.058 & 0.038 &   38 &  0.038 & 0.032 &   25 &  0.054 & 0.041 &  115 \\
  5 & HIP007902       & -0.036 & 0.024 &    8 & -0.087 & 0.023 &    5 & -0.052 & 0.030 &   23 \\
  6 & HIP007918       &  0.031 & 0.026 &   40 &  0.034 & 0.022 &   25 &  0.059 & 0.024 &  115 \\
  7 & HIP011253       &  0.034 & 0.059 &   39 & -0.012 & 0.046 &   25 & -0.013 & 0.043 &  115 \\
  8 & HIP011728       &  0.049 & 0.047 &   16 &  0.033 & 0.027 &   10 &  0.054 & 0.035 &   46 \\
  9 & HIP018413       & -0.045 & 0.027 &   40 & -0.055 & 0.029 &   25 & -0.048 & 0.031 &  115 \\
 10 & HIP021010       & -0.140 & 0.027 &   39 & -0.170 & 0.026 &   25 & -0.110 & 0.033 &  115 \\
 11 & HIP021436       &  0.003 & 0.033 &   39 & -0.002 & 0.026 &   25 &  0.004 & 0.043 &  115 \\
 12 & HIP024813       &  0.106 & 0.026 &    8 &  0.098 & 0.012 &    5 &  0.131 & 0.022 &   23 \\
 13 & HIP029432       & -0.089 & 0.037 &   40 & -0.122 & 0.027 &   25 & -0.104 & 0.031 &  115 \\
 14 & HIP029525       &  0.001 & 0.036 &   40 & -0.040 & 0.033 &   25 & -0.081 & 0.037 &  115 \\
 15 & HIP031965       &  0.026 & 0.034 &   40 &  0.012 & 0.024 &   25 &  0.044 & 0.039 &  115 \\
 16 & HIP035265       &  0.026 & 0.037 &   40 & -0.002 & 0.032 &   26 &  0.006 & 0.031 &  115 \\
 17 & HIP036874       & -0.085 & 0.030 &   16 & -0.168 & 0.034 &   10 & -0.085 & 0.032 &   46 \\
 18 & HIP041484       &  0.019 & 0.025 &   40 &  0.011 & 0.019 &   25 &  0.024 & 0.026 &  115 \\
 19 & HIP041844       & -0.035 & 0.036 &   39 & -0.051 & 0.027 &   25 & -0.015 & 0.041 &  115 \\
 20 & HIP042575       &  0.049 & 0.024 &    8 &  0.033 & 0.025 &    5 &  0.030 & 0.059 &   23 \\
 21 & HIP043557       & -0.048 & 0.037 &    8 & -0.061 & 0.044 &    5 & -0.062 & 0.039 &   23 \\
 22 & HIP043726       &  0.126 & 0.034 &    8 &  0.143 & 0.021 &    5 &  0.126 & 0.038 &   23 \\
 23 & HIP044089       &  0.035 & 0.041 &   40 &  0.029 & 0.027 &   25 &  0.049 & 0.032 &  115 \\
 24 & HIP049756       &  0.037 & 0.043 &   40 &  0.027 & 0.028 &   25 &  0.043 & 0.038 &  115 \\
 25 & HIP050316       &  0.009 & 0.030 &   40 & -0.011 & 0.014 &   24 &  0.028 & 0.040 &  115 \\
 26 & HIP050505       & -0.145 & 0.024 &   39 & -0.187 & 0.027 &   25 & -0.159 & 0.026 &  115 \\
 27 & HIP053721       &  0.003 & 0.027 &   40 &  0.005 & 0.017 &   25 &  0.024 & 0.032 &  115 \\
 28 & HIP056832       &  0.057 & 0.045 &   40 &  0.020 & 0.024 &   25 &  0.037 & 0.038 &  115 \\
 29 & HIP062527       &  0.127 & 0.043 &   40 &  0.170 & 0.025 &   25 &  0.170 & 0.034 &  115 \\
 30 & HIP063636       & -0.031 & 0.049 &   39 & -0.072 & 0.043 &   25 & -0.096 & 0.042 &  115 \\
 31 & HIP064150       &  0.027 & 0.050 &   40 &  0.035 & 0.032 &   25 &  0.049 & 0.040 &  116 \\
 32 & HIP065721       & -0.109 & 0.029 &   40 & -0.159 & 0.018 &   25 & -0.106 & 0.032 &  115 \\
 33 & HIP072043       & -0.047 & 0.035 &   39 & -0.067 & 0.033 &   25 & -0.022 & 0.036 &  115 \\
 34 & HIP076114       & -0.014 & 0.036 &   40 & -0.012 & 0.031 &   25 & -0.001 & 0.035 &  115 \\
 35 & HIP077052       &  0.055 & 0.041 &   40 &  0.041 & 0.029 &   25 &  0.028 & 0.034 &  115 \\
 36 & HIP079672       &  0.034 & 0.026 &    8 &  0.037 & 0.012 &    5 &  0.038 & 0.027 &   23 \\
 37 & HIP085244       &  0.048 & 0.011 &    8 &  0.008 & 0.019 &    5 &  0.027 & 0.034 &   23 \\
 38 & HIP085810       &  0.151 & 0.033 &   40 &  0.177 & 0.030 &   25 &  0.164 & 0.035 &  115 \\
 39 & HIP086193       &  0.043 & 0.035 &   40 &  0.009 & 0.027 &   25 &  0.039 & 0.035 &  115 \\
 40 & HIP088194       & -0.072 & 0.041 &   40 & -0.100 & 0.026 &   25 & -0.057 & 0.032 &  116 \\
 41 & HIP089474       &  0.001 & 0.026 &   40 &  0.003 & 0.019 &   25 &  0.022 & 0.031 &  115 \\
 42 & HIP090355       & -0.149 & 0.029 &   24 & -0.204 & 0.034 &   15 & -0.152 & 0.059 &   69 \\
 43 & HIP094981       &  0.055 & 0.038 &   40 &  0.046 & 0.022 &   25 &  0.023 & 0.033 &  115 \\
 44 & HIP096402       & -0.084 & 0.046 &   16 & -0.163 & 0.018 &   10 & -0.056 & 0.038 &   46 \\
 45 & HIP096895       &  0.107 & 0.032 &    8 &  0.128 & 0.017 &    5 &  0.138 & 0.032 &   23 \\
 46 & HIP096901       &  0.035 & 0.029 &   16 &  0.054 & 0.021 &   10 &  0.075 & 0.033 &   46 \\
 47 & HIP096948       &  0.098 & 0.050 &   24 &  0.101 & 0.036 &   15 &  0.105 & 0.053 &   69 \\
 48 & HIP097336       &  0.089 & 0.034 &   40 &  0.121 & 0.039 &   25 &  0.136 & 0.040 &  115 \\
 49 & HIP097420       &  0.031 & 0.041 &   39 & -0.030 & 0.030 &   25 &  0.001 & 0.040 &  115 \\
 50 & HIP097767       & -0.188 & 0.032 &   40 & -0.291 & 0.035 &   25 & -0.154 & 0.029 &  115 \\
 51 & HIP100963       &  0.004 & 0.047 &   40 & -0.016 & 0.021 &   25 & -0.008 & 0.032 &  115 \\
 52 & HIP102040       & -0.081 & 0.032 &   39 & -0.158 & 0.031 &   25 & -0.099 & 0.027 &  115 \\
 53 & HIP116613       &  0.125 & 0.054 &   40 &  0.095 & 0.020 &   25 &  0.077 & 0.044 &  115 \\
 54 & HIP118162       &  0.066 & 0.038 &   16 &  0.094 & 0.022 &   10 &  0.087 & 0.032 &   46 \\
 55 & TYC2583-01846-2 & -0.035 & 0.035 &   40 & -0.086 & 0.039 &   25 & -0.083 & 0.035 &  115 \\
 56 & TYC2694-00364-1 &  0.012 & 0.040 &   40 & -0.030 & 0.024 &   25 & -0.005 & 0.035 &  115 \\
\hline
\end{tabular}
\end{table*}

\begin{table*}[h]
  \centering 
  \caption{Selected lines from the Gaia-ESO survey  used for the determination of the differental abundances relative to the Sun (excerpt, full version with 267 lines at CDS). The adopted oscillator strength and excitation potential are given for each line, with the reference for loggf.  The Gaia-ESO flags indicate whether the line is recommended (Yes/Undecided/No), based on the quality of the atomic data and of the spectral synthesis for the Sun and Arcturus.}
  \label{t:lines}
\begin{tabular}{llrrcl}
\hline
\hline
wavelength     & element  &  loggf & EP & flags & loggf reference \\
\AA & + ion &  & eV & & \\
\hline
4807.7082 & Fe 1 &  -2.150 &   3.368 & Y- &\citet{2014MNRAS.441.3127R}\\                                                        
4808.1478 & Fe 1 &  -2.690 &   3.252 & YY &\citet{MRW74}\\                                                        
4811.9829 & Ni 1 &  -1.450 &   3.658 & YN &\citet{2003ApJ...584L.107J}\\                                                        
4829.0231 & Ni 1 &  -0.140 &   3.542 & NY &\citet{K08}                     \\                                                        
4844.0135 & Fe 1 &  -2.050 &   3.547 & Y- &\citet{BWL}                    \\                                                        
4848.8826 & Fe 1 &  -3.137 &   2.279 & Y- &\citet{BWL}                   \\                                                        
4855.6729 & Fe 1 &  -1.700 &   3.368 & Y- &\citet{GESHRL14}                \\                                                        
4871.9281 & Fe 1 &  -2.150 &   3.252 & Y- & \citet{BWL}                     \\                                                        
\hline
\end{tabular}
\end{table*}

\section{Additional information}
\label{s:xhip}
The 56 twin candidates are either Hipparcos or Tycho stars.  We have retrieved informations for them in the XHIP catalogue \citep{2012AstL...38..331A} and in Simbad, reported in Table \ref{t:xhip}. All the stars are within 60 pc from the Sun, their spectral type vary from F8 to G6.5. Their B-V colours and M$_v$ absolute magnitudes span a wide range of values. Ages are provided in form of a probable range, sometimes very large due to the lack of constraint on this fundamental property.   

We proceeded like in the previous sections and classified the stars into two categories depending on the similarity of their M$_v$ absolute magnitude and age with the solar ones. Stars with M$_v$ in the range [4.6 -- 5.0] are potentially similar to the Sun (M$_{v_\odot} \simeq 4.80$). If their age interval also includes the age of the Sun ($\sim$4.5 Gyr), then such stars are good twin candidates.  Fifteen stars were found to be similar to the Sun, based on their absolute magnitude and age range.

\begin{table*}[h]
  \centering 
  \caption{Data available in the XHIP catalogue for solar twin candidates. For the two Tycho2 stars, the information was retrieved from Simbad. The last column indicates whether the star has an absolute magnitude and an age compatible to those of the Sun (Yes=A, No=C, Possibly if the information is incomplete=B).}
  \label{t:xhip}
\begin{tabular}{lcclccccccc}
\hline
HIP/TYC  & HD & Dist &  SP  &B-V&M$_v$&U&V&W& age[b-age - B-age]  & age + M$_v$\\
                 &         &  pc  & type &    mag   &    mag & \kms&\kms&\kms&Gyr & \\         
\hline
HIP001813    &HD001832  &       40.26 & F8       & 0.639 & 4.55  & -7.3 & -62.5 & -16.7 & 8.3 [ 6.0 - 10.0 ] & C\\   
HIP004290    &HD005294  &       29.05 & G5       & 0.652 & 5.09  & 33.3 &  -4.5 & -14.5 & 0.2 [     -  4.4 ] & C\\   
HIP007339    &HD009407  &       20.66 & G6.5V    & 0.686 & 4.94  & 50.8 &  -3.9 &   1.1 &                  & B  \\   
HIP007585    &HD009986  &       25.62 & G2V      & 0.648 & 4.73  & -1.1 & -17.2 &  19.0 & 5.4 [ 1.1 -  9.0 ] & A\\   
HIP007902    &HD010145  &       37.22 & G5V      & 0.691 & 4.85  &113.8 & -63.7 & -20.7 & 7.3 [ 2.9 - 11.3 ] & A \\   
HIP007918    &HD010307  &       12.74 & G1V      & 0.618 & 4.43  &-38.6 & -31.2 &  -0.1 & 6.1 [ 3.5 -  8.2 ] & C\\   
HIP011253    &HD014874  &       60.29 & G0V      & 0.665 & 4.26  & 16.0 & -20.7 & -47.9 &10.0 [ 8.3 - 11.7 ] & C \\   
HIP011728    &HD015632  &       39.77 & G0       & 0.666 & 5.03  &-24.2 &  -8.6 &  16.3 & 0.8 [     -  6.7 ] & C\\   
HIP018413    &HD024409  &       21.98 & G3V      & 0.698 & 4.82  & 34.5 &  15.3 &  -5.9 &13.2 [ 9.4 - 16.3 ] & C\\   
HIP021010    &HD028447  &       38.54 & G5       & 0.722 & 3.58  &-31.3 &  -5.5 &  18.4 & 7.3 [ 6.7 -  7.8 ] & C \\   
HIP021436    &HD029150  &       32.57 & G5       & 0.685 & 5.02  &  6.4 &  -7.3 &   0.2 & 2.6 [     -  8.6 ]  & C\\   
HIP024813    &HD034411  &       12.63 & G1V      & 0.630 & 4.18  &-75.7 & -35.1 &   4.4 & 5.2 [ 3.8 -  6.5 ] & C \\   
HIP029432    &HD042618  &       23.50 & G3V      & 0.642 & 4.99  & 62.1 & -12.6 &  11.2 & 0.2 [     -  4.3 ] & C\\   
HIP029525    &HD042807  &       17.95 & G5V      & 0.663 & 5.16  &  3.0 & -25.9 &  -6.6 &                   & C \\   
HIP031965    &HD047309  &       40.25 & G0       & 0.672 & 4.58  &-45.6 & -41.4 &  -8.3 &12.0 [ 8.4 - 14.1 ] & C \\   
HIP035265    &HD056124  &       27.08 & G0       & 0.631 & 4.77  &-24.5 & -20.4 &  -6.5 & 0.6 [     -  5.1 ] & A \\   
HIP036874    &HD060298  &       39.06 & G2V      & 0.642 & 4.41  &140.0 & -35.6 & -48.2 & 8.7 [ 7.1 - 12.0 ] & C \\   
HIP041484    &HD071148  &       22.25 & G1V      & 0.624 & 4.58  & 19.9 & -38.9 & -22.7 & 7.6 [ 4.0 -  9.8 ] & C \\   
HIP041844    &HD071881  &       41.25 & G1V      & 0.630 & 4.36  &-33.3 & -60.9 &  -4.6 & 6.8 [ 4.8 -  8.4 ] & C\\   
HIP042575    &HD073393  &       39.86 & G3V      & 0.675 & 5.00  &-74.2 & -54.4 & -12.6 & 7.2 [ 0.8 - 12.7 ] & A \\   
HIP043557    &HD075767  &       24.00 & G0V      & 0.640 & 4.67  & 17.9 & -26.7 &   6.1 & 9.8 [ 5.5 - 11.0 ] & C\\   
HIP043726    &HD076151  &       17.38 & G3V      & 0.661 & 4.81  &-40.5 & -19.8 & -11.7 & 7.1 [ 3.1 - 11.0 ] & A\\   
HIP044089    &HD076752  &       38.50 & G2V      & 0.680 & 4.54  &  7.1 & -20.2 & -20.9 & 8.5 [ 6.1 - 10.4 ] & C \\   
HIP049756    &HD088072  &       34.91 & G3V      & 0.647 & 4.83  &-19.6 &   4.5 & -32.4 & 3.7 [     -  7.9 ] & A \\   
HIP050316    &HD088986  &       33.18 & G2V      & 0.635 & 3.86  &-19.1 & -21.5 &  17.5 & 6.5 [ 4.2 -  7.4 ] & C \\   
HIP050505    &HD089269  &       20.24 & G6V      & 0.653 & 5.13  & 12.6 & -27.5 &   1.7 & 7.7 [ 1.3 - 13.8 ] & C \\   
HIP053721    &HD095128  &       14.06 & G1V      & 0.624 & 4.29  &-24.2 &  -2.5 &   0.8 & 6.8 [ 4.9 -  8.1 ] & C \\   
HIP056832    &HD101242  &       35.04 & G6       & 0.710 & 4.88  &-41.1 & -48.9 &  -1.4 & 2.7 [     -  7.1 ] & A \\   
HIP062527    &HD111513  &       38.08 & G1V      & 0.633 & 4.46  &-85.8 & -51.4 & -28.7 & 6.9 [ 4.9 -  9.0 ] & C \\   
HIP063636    &HD113319  &       31.09 & G4V      & 0.655 & 5.05  &-27.5 &  -9.8 &   7.0 & 7.0 [     - 12.2 ]  & C \\   
HIP064150    &HD114174  &       26.12 & G3IV     & 0.667 & 4.69  & 57.2 & -64.8 &  -9.5 & 8.8 [ 6.1 - 11.3 ] & C \\   
HIP065721    &HD117176  &       17.99 & G5V      & 0.714 & 3.70  & 12.9 & -51.4 &  -4.3 & 8.0 [ 7.5 -  8.6 ] & C \\   
HIP072043    &HD129814  &       41.09 & G5V      & 0.636 & 4.45  & 15.1 & -30.2 &   5.8 & 8.4 [ 6.1 - 10.2 ] & C \\   
HIP076114    &HD138573  &       30.24 & G5IV-V   & 0.656 & 4.82  &-36.6 &   8.8 & -19.2 & 7.1 [ 2.6 - 11.3 ] & A \\   
HIP077052    &HD140538  &       14.66 & G5V      & 0.684 & 5.03  & 18.1 &  -7.3 &  10.7 & 2.8 [     -  8.3 ] & C \\   
HIP079672    &HD146233  &       13.90 & G2V      & 0.652 & 4.77  & 27.1 & -14.4 & -22.1 & 2.4 [     -  6.8 ] & A \\   
HIP085244    &HD158222  &       41.54 & G0       & 0.667 & 4.73  &-18.4 & -10.9 &  -5.2 &10.2 [ 6.9 - 12.4 ] & C \\   
HIP085810    &HD159222  &       23.92 & G1V      & 0.639 & 4.63  &-31.0 & -49.9 &  -1.9 & 4.3 [ 1.3 -  7.8 ] & A \\   
HIP086193    &HD159909  &       35.18 & G5       & 0.693 & 4.55  &-59.0 & -54.4 &  -7.7 &13.0 [10.8 - 15.3 ] & C\\   
HIP088194    &HD164595  &       28.35 & G2V      & 0.635 & 4.81  &-17.7 &   2.3 &  24.0 &10.7 [ 6.6 - 13.6 ] & C \\   
HIP089474    &HD168009  &       22.82 & G1V      & 0.641 & 4.51  & -4.5 & -62.1 & -22.7 & 9.1 [ 7.4 - 12.6 ] & C \\   
HIP090355    &HD169822  &       28.84 & G6V      & 0.699 & 5.53  & 28.5 & -65.3 &  -6.6 &                    & C \\   
HIP094981    &HD181655  &       25.39 & G5V      & 0.676 & 4.27  & 22.4 &  -5.7 &  -2.5 &10.4 [ 8.7 - 11.9 ] & C \\   
HIP096402    &HD184768  &       38.56 & G5 V     & 0.675 & 4.62  & 24.8 & -61.4 & -27.9 &14.4 [11.9 - 16.7 ] & C\\   
HIP096895    &HD186408  &       21.08 & G1.5V    & 0.643 & 4.37  & 17.4 & -30.5 &  -0.5 & 7.1 [ 5.3 -  9.7 ] & C \\   
HIP096901    &HD186427  &       21.21 & G3V      & 0.661 & 4.62  & 17.2 & -30.5 &  -2.1 & 7.6 [ 4.7 -  9.5 ] & C \\   
HIP096948    &HD186104  &       41.19 & G5 V     & 0.664 & 4.57  &-53.7 & -39.8 & -21.2 & 8.2 [ 5.7 - 10.2 ] & C \\   
HIP097336    &HD187123  &       48.22 & G2V      & 0.661 & 4.41  &  3.1 & -15.4 & -43.8 & 5.0 [ 3.0 -  6.8 ] & C \\   
HIP097420    &HD187237  &       26.24 & G2IV-V   & 0.660 & 4.78  &-35.5 & -19.6 &  14.2 & 8.6 [ 4.5 - 12.1 ] & A \\   
HIP097767    &HD187923  &       26.61 & G0   V   & 0.642 & 4.03  & 33.2 & -51.1 &  19.3 &10.9 [10.0 - 11.8 ] & C \\   
HIP100963    &HD195034  &       28.22 & G5       & 0.642 & 4.84  & 24.1 & -15.9 & -15.5 & 7.6 [ 3.0 - 11.6 ] & A \\   
HIP102040    &HD197076  &       20.94 & G1V      & 0.611 & 4.82  &-42.9 & -15.2 &  16.3 & 5.5 [ 0.2 -  9.3 ] & A \\   
HIP116613    &HD222143  &       23.33 & G3V      & 0.665 & 4.74  &-34.1 & -15.9 & -12.3 & 6.2 [ 2.0 -  9.7 ] & A \\   
HIP118162    &HD224465  &       24.52 & G4V      & 0.694 & 4.77  & -4.9 &   8.8 &  27.5 & 6.3 [ 1.5 -  9.5 ] & A \\   
 TYC2583-01846-2 &  HD146362 & 22.22 & G1V & 0.621  & 5.00 & -9& -33 &  7 &   & B \\
 TYC2694-00364-1 &    HD197310 &  &                 & 0.60   &         &     &         &     &    & B \\
\hline
\end{tabular}
\end{table*}

We have also searched in Simbad whether the twin candidates have a detected extrasolar planet and found 4 such stars : HIP053721, HIP065721, HIP096901, HIP097336, none of them being a good solar twin.

\section{Discussion}
\label{s:discuss} 
In each of the previous sections, we have classified 56 twin candidates according to their similarity with the Sun using different properties :  global spectrum, atmospheric parameters, Li content, abundances of 8 chemical elements, absolute magnitude, age, presence of a planet. According to the TGMET method, 6 stars were found to have a spectrum very similar to that of the Sun  : HIP079672, HIP018413, HIP076114, HIP089474, HIP118162, HIP041484. 
Three stars were found to have solar atmospheric parameters : HIP076114, HIP085244 and HIP088194, two of which with also the same Li deficiency : HIP076114,  HIP088194.  
Seven stars were found to have the solar abundance pattern for 8 chemical elements : HIP021436, HIP035265, HIP076114, HIP085244, HIP089474, HIP100963, TYC2694-00364-1. Table \ref{t:summary} summarizes the categories A, B and C that we gave to each target according to its similarity to the Sun in atmospheric parameters, abundances, Li content, absolute magnitude and age.

\begin{table*}[h]
  \centering 
  \caption{Summary of the classification of solar twin candidates after comparison to the Sun based on atmospheric parameters, abundances of 8 elements, Li content, absolute magnitude and age. The best solar twins are sorted at the top of the table. The columns AP (atmospheric parameters), Li and Mv+age report the classification (A, B, C) from Tables \ref{t:ap} and \ref{t:xhip}. The column Ab. indicates whether the differential abundances in Table \ref{t:fena},\ref{t:sicati}, \ref{t:crmnni} agree with the solar value within 0.05 dex (A) or within 0.10 dex (B), or differ (C). For the 19 stars considered as good solar twins (AP and Ab. equal A or B), the last column indicates studies where that star was previously identified  as solar twin : [1] = \cite{2004A&A...418.1089S}, [2] = \cite{2006ApJ...641L.133M}, [3] =  \cite{2009A&A...508L..17R}, [4] = \cite{2009PASJ...61..471T}, [5] = \cite{2012MNRAS.426..484D}, [6] = \cite{2014A&A...563A..52P}, [7] = \cite{2014A&A...572A..48R}, [8] = \cite{2015A&A...574A.124D}}
  \label{t:summary}
\begin{tabular}{llccccl}
\hline
HIP/TYC  & HD &  AP & Ab. & Li & Mv+age& previous studies \\
\hline
\multicolumn{7}{c}{\bf solar twins } \\
HIP076114       & HD138573 & A & A & A & A &  [5], [6], [7]\\
HIP085244       & HD158222 & A & A & B & C & [6]\\
HIP088194       & HD164595 & A & B & A & C & [6]\\
HIP021436       & HD029150 & B & A & A & C & [8]\\
HIP035265       & HD056124 & B & A & C & A & [2]\\
HIP100963       & HD195034 & B & A & C & A &  [3], [4]\\
TYC2694-00364-1 & HD197310 & B & A & C & B &  \\
HIP042575       & HD073393 & B & B & A & A & \\
HIP056832       & HD101242 & B & B & A & A & \\
HIP118162       & HD224465 & B & B & A & A & \\
HIP011728       & HD015632 & B & B & A & C &   [6], [8]\\
HIP018413       & HD024409 & B & B & A & C & \\
HIP049756       & HD088072 & B & B & B & A &   [1], [6], [7], [8]\\
HIP079672       & HD146233 & B & B & B & A & [1], [2], [3], [4], [5], [6], [7], [8]\\
HIP043557       & HD075767 & B & B & B & C &   [8]\\
HIP007585       & HD009986 & B & B & C & A &  [2], [6], [7], [8]\\
HIP097420       & HD187237 & B & B & C & A &   [6]\\
HIP041484       & HD071148 & B & B & C & C &  [1], [2] \\
HIP094981       & HD181655 & B & B & C & C & \\
\hline
\multicolumn{7}{c}{ \bf other stars } \\
HIP011253       & HD014874 & B & C & B & C &\\
HIP029432       & HD042618 & B & C & B & C &\\
HIP102040       & HD197076 & B & C & C & A &\\
HIP007339       & HD009407 & C & A & A & B &\\
HIP089474       & HD168009 & C & A & A & C &\\
HIP001813       & HD001832 & C & B & A & C &\\
HIP031965       & HD047309 & C & B & A & C &\\
HIP044089       & HD076752 & C & B & A & C &\\
HIP064150       & HD114174 & C & B & A & C &\\
HIP072043       & HD129814 & C & B & A & C &\\
HIP086193       & HD159909 & C & B & A & C &\\
HIP041844       & HD071881 & C & B & B & C &\\
HIP077052       & HD140538 & C & B & B & C &\\
HIP096901       & HD186427 & C & B & B & C &star with planet\\
HIP007918       & HD010307 & C & B & C & C &\\
HIP050316       & HD088986 & C & B & C & C &\\
HIP053721       & HD095128 & C & B & C & C & star with planet\\
HIP007902       & HD010145 & C & C & A & A &\\
HIP036874       & HD060298 & C & C & A & C &\\
HIP050505       & HD089269 & C & C & A & C &\\
HIP090355       & HD169822 & C & C & A & C &\\
HIP096402       & HD184768 & C & C & A & C &\\
HIP096895       & HD186408 & C & C & A & C &\\
HIP097336       & HD187123 & C & C & A & C &star with planet\\
HIP097767       & HD187923 & C & C & A & C &\\
HIP062527       & HD111513 & C & C & B & C &\\
HIP096948       & HD186104 & C & C & B & C &\\
HIP043726       & HD076151 & C & C & C & A &\\
HIP085810       & HD159222 & C & C & C & A &\\
HIP116613       & HD222143 & C & C & C & A &\\
TYC2583-01846-2 & HD146362 & C & C & C & B &\\
HIP004290       & HD005294 & C & C & C & C &\\
HIP021010       & HD028447 & C & C & C & C &\\
HIP024813       & HD034411 & C & C & C & C &\\
HIP029525       & HD042807 & C & C & C & C &\\
HIP063636       & HD113319 & C & C & C & C &\\
HIP065721       & HD117176 & C & C & C & C & star with planet\\
\hline
\end{tabular}
\end{table*}

 It is remarkable to find a star similar to the Sun in all these properties : HIP076114 (HD138573), supported also by the absolute magnitude and most probable age range. This star is also in the list of best twins by \cite{2012MNRAS.426..484D} although they classifiy it after HIP079672 (18 Sco) \citep[see also][]{2015A&A...574A.124D}. HIP076114 is also part of the survey of solar twin stars within 50 parsecs of the Sun by \cite{2014A&A...563A..52P}, as well as in the Solar Twin Planet Search by \cite{2014A&A...572A..48R}, but not mentioned as a remarkable twin in these two papers.  It is worth noting that the galactic velocity of that star differs by $\sim$40 \kms\ from that of the Sun, suggesting that it is not a solar  sibling. It is however an excellent star to use for calibrations or to search for exoplanets.

HIP085244 is our second best twin candidate, with solar atmospheric parameters and chemical pattern but it is classified B for the Li, with a slightly more enhanced content.  According to XHIP this star is supposed to be older than the Sun. HIP085244 is also part of the survey of solar twin stars within 50 parsecs of the Sun by \cite{2014A&A...563A..52P}.

HIP088194 has solar atmospheric parameters and is deficient in Li, but it is slightly more metal-poor than the Sun. This is clearly visible in the Figures \ref{f:fe} and \ref{f:el} (star 40). It is older than the Sun according to XHIP, but of the same age according to \cite{2014A&A...563A..52P}. It is one of the new candidates considered as excellent in that work. 

HIP021436 has the solar chemical pattern, and also the same Li content but its effective temperature is lower by $\sim$60 K than that of the Sun. Despite us classifying it  B for the atmospheric parameters, it still falls  into the solar twin category according to the criterion defined by \cite{2014A&A...567L...3M}. Our \teff\ determination (5715$\pm$10 K) is in very good agreement with the 7 values listed in the PASTEL catalog \citep{2010A&A...515A.111S} ranging from 5675 to 5748 K. It is worth noting that this star has a galactic velocity compatible with that of the Sun within 10 \kms. Its age reported in XHIP is not well constrained, so possibly compatible with that of the Sun. It is thus also a very good solar sibling candidate.

HIP035265 has the solar chemical pattern, an absolute magnitude and age compatible with the Sun, but it is hotter by $\sim$80 K and with a strong Li feature. It also has a galactic velocity different from the Sun. This is thus neither a perfect solar twin or sibling.

HIP100963 has the solar chemical pattern, but it is hotter than the Sun (\teff = 5821$\pm$6 K) and has a pronounced Li feature. It was previously identified as solar twin by \cite{2009A&A...508L..17R} who determined \teff = 5815 K and \feh=+0.018$\pm$0.019 in very good agreement with our determinations (we determined \feh=+0.011$\pm$0.034). Considering this star as an excellent solar twin, \cite{2009PASJ...61..471T} made an extensive study at very high resolution (R=90\,000) and high S/N (500-1000) and found it hotter than the Sun by 23 K, with a difference of iron abundance of 0.004 and a higher Li content by a factor of $\sim$56 as compared to Sun, also in good agreement with our findings. 

TYC2694-00364-1 has been poorly studied as an individual star until now. It essentially differs from the Sun by its hotter temperature (\teff = 5842$\pm$1 K) and a strong Li feature. 

HIP042575, HIP056832 and  HIP118162 are our next good solar twins, never identified as such before. Although slightly metal-rich,  their abundances differ by less than 0.1 dex from those of the Sun. They exhibit the solar Li deficiency and have an age and absolute magnitude compatible with those of the Sun. Next is HIP011728, also more metal-rich but previously identified as solar twin by \cite{2014A&A...563A..52P} and \cite{2015A&A...574A.124D}.  Its absolute magnitude is reported as Mv=5.03 in XHIP, higher than that of the Sun. Then we have HIP018413, a new solar twin, slightly colder, more  metal-poor and older than the Sun, which was one of the closest twins according to TGMET. HIP049756 was already identified as a solar twin in several previous studies. We classified it B for the atmospheric parameters and abundances because of [M/H] and [Ca/H] being slightly higher than our limits, 0.05 dex and 0.058 dex respectively, while all the other values are solar. Its main difference to the Sun is a higher lithium content.

The most studied  HIP079672 (18 Sco) is definitively not a perfect solar twin, according to our ranking at the 14th position in Table \ref{t:summary}. We find a hotter temperature and enhanced abundances relative to the Sun, including the lithium, thus confirming the findings of \cite{2014ApJ...791...14M}.

The next four solar twins are already known. They have all a higher lithium content. The last one of the list, HIP094981, is unknown. It  is colder and more metal rich but still a solar twin according to the criterion of \cite{2014A&A...567L...3M}. It is also more luminous (Mv=4.27) and much older than the Sun (10.4 Gyr) with a strong Li feature.

The second part of Table \ref{t:summary} lists the other stars which were found to differ from the Sun, either in their atmospheric parameters or in their abundances.  It is however worth to note HIP007339 and  HIP089474 which have the solar chemical pattern and lithium deficiency. HIP007339  is one of the coldest star of our sample (\teff=5630$\pm$12 K) while HIP089474 differs from the Sun by its lower surface gravity (\logg = 4.29$\pm$0.01). Both stars are not however good solar sibling candidates, having a galactic velocity significantly different from the Sun, with a radial velocity component U=50.8 \kms for HIP007339 and a rotational component V= -62.1 \kms for HIP089474. HIP089474 is also old (9.1 Gyr), which interestingly makes it a possible member of the thick disk.

 \section{Conclusion}
 In this work, we have selected 56 solar twin candidates, among 1165 tested stars, based on the similarity of their spectrum to that of the Sun. We have determined their atmospheric parameters and abundances, and examined their Li content to further study their similarity to the Sun.
 
 A strength of our study is that we have used as reference a large number of solar spectra (18), representive of the range of observing conditions, and from which we have selected the best lines to be analyzed for abundance measurements by synthesis. Several spectra were available for the majority of the selected targets, up to 5 (225 spectra in total). The analysis was performed differentially to the solar spectra, on a line by line basis for the selected set of lines. This led to a very good internal precision. Our atmospheric parameters and abundances proved to also be accurate, in excellent agreement with other studies made at higher resolution and S/N.  This demonstrates that the \elodie\ archive is a very good resource to make comprehensive studies of solar type stars.
 
 We found that the best solar twin in our sample is HIP076114 (HD138573), dethroning previous candidates (HIP079672/18 Sco, HIP100963). All the other stars slightly differ from the Sun in one or another property. We list 19 solar twins which have \teff, \logg\ and abundances differing from those of the Sun by less than 100 K and 0.1 dex respectively.  
 
 We found a good solar sibling candidate : HIP021436 which would be worth studying further.
 
 \begin{acknowledgements}
CS would like to dedicate this paper to  the memory of Giusa Cayrel de Strobel, an inspiration for a younger generation of astronomers in
the quest for solar twins and many other topics.   DM acknowledges financial support from the Irakian ministry of higher education and research. We acknowledge constructive comments and suggestions by the anonymous referee. We  used  the CDS-SIMBAD and NASA-ADS databases, and the VizieR service at CDS.
\end{acknowledgements}

\end{document}